\documentclass[twocolumn]{aastex61}

\usepackage[english]{babel}
\usepackage[utf8]{inputenc}
\usepackage{amsmath}
\usepackage{graphicx}
\usepackage{physics}
\usepackage{float}

\received{}
\revised{\today}
\accepted{}
\submitjournal{}

\shorttitle{High-Resolution Spectra Interpreted with Doppler Shifts from Three-Dimensional Models}
\shortauthors{Flowers et al.}

\begin{document}

\title{The high-resolution transmission spectrum of HD 189733\lowercase{b} interpreted with atmospheric Doppler shifts from three-dimensional general circulation models}

\author[0000-0001-8045-1765]{Erin Flowers}
\affil{Department of Astrophysical Sciences, Princeton University, 4 Ivy Lane, Princeton, NJ 08540, USA}
\affil{Department of Astronomy, Columbia University, 538 West 120th Street, New York, NY 10027, USA}

\author[0000-0002-7704-0153]{Matteo Brogi}
\affil{Department of Physics, University of Warwick, Coventry CV4 7AL, UK}
\affil{INAF - Osservatorio Astrofisico di Torino, Via Osservatorio 20, 10025, Pino Torinese, Italy}
\affil{Centre for Exoplanets and Habitability, University of Warwick, Gibbet Hill Road, Coventry CV4 7AL, UK}

\author[0000-0003-3963-9672]{Emily Rauscher}
\affil{Department of Astronomy, University of Michigan, 1085
 S. University Avenue, Ann Arbor, MI 48109, USA}

\author[0000-0002-1337-9051]{Eliza M.-R.\ Kempton}
\affil{Department of Astronomy, University of Maryland, College Park, MD 20742, USA}
\affil{Department of Physics, Grinnell College, 1116 8th Avenue, Grinnell, IA 50112, USA}

\author[0000-0003-3891-7554]{Andrea Chiavassa}
\affil{Universit\'e C\^ote d'Azur, Observatoire de la C\^ote d'Azur, CNRS, Lagrange, CS 34229, Nice, France}

\begin{abstract}
The signature of wind patterns caused by the interplay of rotation and energy redistribution in hot Jupiters is detectable at high spectral resolution, yet no direct comparison has been attempted between predictions from general circulation models (GCMs) and observed high-resolution spectra. We present the first of such comparisons on near-infrared transmission spectra of the hot Jupiter HD 189733b. Exploring twelve rotation rates and two chemical regimes, we have created model spectra from 3-D GCMs and cross-correlated them with the observed spectra. Comparing our models against those of HD 189733b, we obtain three key results: (1) we confirm CO and H$_2$O in the planet's atmosphere at a detection significance of 8.2$\sigma$; (2) we recover the signature of $\sim$km/s day-to-night winds at $\sim$mbar pressures; and (3) we constrain the rotation period of the planet to between 1.2 and 4.69 days (synchronous rotation (2.2 days) remains consistent with existing observations). Our results do not suffer from the shortcomings of 1-D models as cross correlation templates -- mainly that these models tend to over-constrain the slower rotation rates and show evidence for anomalous blue shifts. Our 3-D models instead match the observed line-of-sight velocity of this planet by self-consistently including the effects of high-altitude day-to-night winds. Overall, we find a high degree of consistency between HD 189733b observations and our GCM-based spectra, implying that the physics and chemistry are adequately described in current 3-D forward models for the purpose of interpreting observations at high spectral resolution.
\end{abstract}

\keywords{hydrodynamics---radiative transfer---planets and satellites: atmospheres---planets and satellites: gaseous planets}

\section{Introduction}

Within the last decade, the characterization of exoplanet atmospheres has been significantly enhanced by the development of ground-based measurements using high-resolution spectroscopy (HRS, $R \sim 25,000-100,000$). This powerful tool separates the planet and stellar spectra via the Doppler shift induced by the planet's orbital motion (for close-in planets), or through spatial isolation enabled by high contrast imaging (for wide-orbit planets); see \citet{Birkby2018} for a recent review. In addition to enabling chemical and physical characterization of exoplanet atmospheres, those measurements at the highest resolution ($R\sim 100,000$) are also sensitive to distortions in the planetary lines due to high-speed winds and/or global rotation \citep{Miller-RicciKempton2012CONSTRAININGTRANSIT,Showman2013DopplerJupiters,Kempton2014HighModels,Rauscher2014THEJUPITERS,Zhang2017Emission}.
In the very first demonstration of the HRS method, \citet{Snellen2010TheHD209458b} marginally detected an overall blue-shift of $-2\pm 1$\,km\,s$^{-1}$ in the VLT/CRIRES transmission spectrum of the hot Jupiter HD 209458b, which was tentatively interpreted as due to high-altitude winds flowing across the planet terminator, from day to night. A few years later, line broadening in VLT/CRIRES spectra of wide-orbit companions was used to infer the fast rotational velocity ($v_{rot} = 25 \pm 3$ km s$^{-1}$) of $\beta$ Pictoris b \citep{Snellen2014FastB} and the much slower rotation ($v_{rot} = 5 \pm 1$ km s$^{-1}$) of GQ Lupi b \citep{Schwarz2016TheConfiguration}.

The close-in hot Jupiters are expected to have been tidally locked into synchronous rotation, meaning the rotation and orbital periods are equal, with characteristic values of a few days. This corresponds to expected rotational velocities of a few km s$^{-1}$, which is the same order of magnitude as our expectations for wind speeds on these planets \citep[e.g.,][]{Showman2002AtmosphericPlanets}. Beyond the tentative wind detection reported in \citet{Snellen2010TheHD209458b}, \citet{Louden2016NaABSORPTION} studied the transmission spectrum of another hot Jupiter (HD 189733b) and attempted to model the radial velocities of the leading and trailing parts of the planet terminator separately. They detected an overall blue shift of $-1.9^{+0.7}_{-0.6}$ km s$^{-1}$ and a significant velocity difference between the trailing and leading limbs. Under the hypothesis of rigid-body rotation, the latter was interpreted as indicative of equatorial wind speeds exceeding the pure rotational regime. Atmospheric motion of this planet was also independently detected in the infrared by \citet{Brogi2016RotationSpectroscopy} via VLT/CRIRES observations. The strong H$_2$O + CO absorption (7.6$\sigma$), was found to be broadened significantly and well modelled with a global blueshift of $-1.7^{+1.1}_{-1.2}$ km s$^{-1}$ and a planet's rotational velocity compatible with synchronous rotation, while rotational periods faster than one day were confidently ($>3\sigma$) excluded. These two studies apply a very different parametrization of the velocity field of the planet terminator, however they both draw conclusions about planetary winds by interpreting measured radial velocities under the simplified hypothesis of rigid rotation. Given their exploratory nature, there is no attempt of modelling the interplay between rotation and development of large scale atmospheric dynamics from first principles (e.g. by solving the relevant physical equations).

One key aspect of the HRS method is that it extracts the planetary signal via cross correlation of the data with atmospheric models, with the strength of the cross-correlation function indicating the degree of match between model and data. All previous inferences on planetary winds and rotation were obtained by cross-correlating data with spectra from one-dimensional models, with the planetary lines allowed to broaden and/or shift based on some simple parameters (e.g. a simple Doppler shift for day-to-night winds and/or Doppler broadening from solid body rotational velocities). However, treating these shifts and broadenings as effects that can be independently parameterized oversimplifies the complex physical reality of the atmosphere, where the planet's wind pattern is fundamentally shaped by the planet's rotation rate; the Coriolis force is one of the main balancing forces in global atmospheric dynamics. The planet's true atmospheric Doppler signature will be from a complex combination of the multi-dimensional wind field and global rotation rate. 

For the first time, we present an analysis of high-resolution data in which we use modeled spectra for the cross-correlation, where the spectra come from the results of a suite of three-dimensional general circulation models (GCMs), directly predicting the atmospheric velocity field for different rotation rates. We successfully use these template spectra to constrain winds and rotation of the planet HD 189733b, reanalyzing the VLT/CRIRES high-resolution transmission spectra presented in \citet{Brogi2016RotationSpectroscopy}.

In Section \ref{sec:model} we describe the setup of the GCM used to predict the three dimensional atmospheric structure and the radiative transfer model used to calculate transmission spectra with Doppler shifts and broadenings. We then present the results from each modeling component in Sections \ref{sec:gcmres} and \ref{sec:transspecres}. In Section \ref{sec:obs} we then describe the observations of HD 189733b that are used in our analysis, cross-correlating the data with models in Section \ref{sec:ccf}. Finally, we discuss our results and summarize the most salient points in Section \ref{sec:dis}.

\section{Computational Modeling Methods} \label{sec:model}
In this study we use a General Circulation Model (GCM) that solves a standard set of simplified fluid dynamics and radiative transfer equations to simulate the three-dimensional atmospheric temperature and wind structure for the hot Jupiter HD 189733b. Specifically, we use the GCM from \citet{Rauscher2012ATRANSFER}, which solves the primitive equations of meteorology and uses two-stream double-gray radiative transfer. We calculate a set of twelve models with rotation periods ranging from 1.08 to 18.1 days, including one at the synchronous rotation period of 2.22 days (see Table \ref{tab:modelpar}). This is a significant expansion of our initial study of three rotation cases (two non-synchronous and the synchronous case) of HD~189733b \citep{Rauscher2014THEJUPITERS}, and we have also updated the planet's radius and gravity, based on recent interferometric observations that better measure the star's radius \citep{Boyajian2014StellarDwarfs}. The system parameters used are shown in Table \ref{tab:planetstarpar}. As in \citet{Rauscher2014THEJUPITERS}, we use 45 vertical levels, evenly spaced in log pressure from 100 bar to 10 microbar and a horizontal resolution of T31, corresponding to $\sim$ 4 degrees in latitude and longitude, which we showed as sufficient to resolve the atmospheric flow. The resolution element is much smaller than the Rossby deformation radius of even our most quickly rotating model. We initialize each model from rest (zero winds) and run for 2000 planet orbits, by which point all of the higher (observable) levels of the atmosphere reach a steady state.

%rotation rates
\startlongtable
\begin{deluxetable}{cccc}
\tablewidth{1.0\columnwidth}
\tablecaption{Suite of 3D Circulation Models \label{tab:modelpar}}
\tablehead{
\colhead{Rotation} & \colhead{$v_{\mathrm{rotation}}$} & \colhead{Max. $v_{\mathrm{wind}}$ at} & \colhead{Max. $v_{\mathrm{wind}}$ at} \\
\colhead{Period} & \colhead{at equator} & \colhead{IR photosphere} & \colhead{0.1 mbar} \\
\colhead{(days)} & \colhead{(km/s)} & \colhead{(km/s)} & \colhead{(km/s)} }
\startdata
18.1 & 0.349 & 4.411 & 5.722\\
7.45 & 0.849 & 5.165 & 7.077\\
4.69 & 1.349 & 7.953 & 9.491\\
3.42 & 1.849 & 6.363 & 6.926\\
2.69 & 2.349 & 5.342 & 6.147\\
\textbf{2.22} & \textbf{2.849} & \textbf{4.303} & \textbf{5.148}\\
1.89 & 3.349 & 4.081 & 4.857\\
1.64 & 3.849 & 3.376 & 4.058\\
1.45 & 4.349 & 3.159 & 4.550\\
1.30 & 4.849 & 2.756 & 4.458\\
1.18 & 5.349 & 2.474 & 3.744\\
1.08 & 5.849 & 1.154 & 2.857\\
\enddata
\tablenotetext{}{Synchronous rotation case in bold. Wind speeds are calculated within the rotating frame of the planet and are from the infrared photosphere (130 mbar) and 0.1 mbar pressure levels, representative of the heights probed by emission and transmission measurements, respectively.} 
\end{deluxetable}

The output of the GCM is then post-processed by a radiative transfer code to calculate the wavelength-dependent attenuation of stellar light through the planetary atmosphere. This radiative transfer code is based on the 1-D model originally presented in \citet{Miller-Ricci2009TheAtmospheres}, which was later made publicly available as the \texttt{Exo-Transmit} package \citep{Kempton2017ExoTransmit}. The proprietary version of \texttt{Exo-Transmit} used in this study (see \citealt{Miller-RicciKempton2012ConstrainingTransitb}) further accounts for the local temperature, pressure, composition, and line-of-sight wind speed within each grid cell of the 3-D model to compute the gas opacity, rather than assuming a radially isotropic T-P profile. As input into the radiative transfer code, the pressure-based vertical grid from the GCM is interpolated onto one that is spaced equally in altitude, which allows us to more easily strike straight-line rays through the atmosphere. Note that this formalism works well for the upper atmosphere (pressures less than ~1 mbar) that we are mainly probing, where refraction should have a negligible impact on hot Jupiter atmospheres (e.g. \citealt{Betremieux2014Clouds}). We further neglect the effects of aerosols, which are not included in our current model. Over the very limited wavelength range of our current model-to-observation comparison, aerosols should act as gray absorbers, which would primarily impact the perceived continuum level of the transmission spectra. As the continuum information is lost in the observational data reduction process, we conclude that our current analysis is not sensitive to the presence of aerosols and we are therefore justified in neglecting their effects. The radiative transfer equation is then solved for the case of pure absorption
\begin{equation}\label{eq:inten}
I(\lambda) = I_0e^{-\tau(\lambda)}
\end{equation}
\noindent
where $I_0$ is the incident stellar intensity and $\tau$ is the line-of-sight ``slant" optical depth as a function of wavelength. The latter is calculated according to 
\begin{equation}\label{eq:taulos}
\tau(\lambda) = \int \kappa(\lambda) ds.
\end{equation}
In Equation~\ref{eq:taulos}, $\kappa$ is the local gas opacity in a given grid cell of the 3-D atmosphere, and $ds$ is the line-of-sight path length traveled through that grid cell. The optical depth $\tau$ is calculated along a total of $2 \times \textrm{NALT} \times \textrm{NLAT}$ individual rays, where $\textrm{NALT}$ is the number of vertical levels, and $\textrm{NLAT}$ is the number of latitude grid cells in the 3-D model. The factor of two comes from integrating the light propagating through the planetary atmosphere on both the eastern and western limbs.

Doppler shifts resulting from winds and rotation need to be accounted for because these effects are often larger than both the natural width of individual molecular lines and the resolution elements of a high resolution spectrograph (e.g. CRIRES with $R \sim 100,000$). For this reason, the local opacity in each grid cell is Doppler shifted according to the line-of-sight gas velocity ($v_{LOS}$), given by
\begin{equation}\label{eq:los_eq}
\begin{aligned}
v_{LOS} = -(u\sin\theta + v\cos\theta\sin\phi
+ (R_p + z) \Omega\sin\theta)
\end{aligned}
\end{equation}
\noindent
The first part of the equation calculates the line of sight velocities caused by the winds ($u$ and $v$ being the east-west and north-south components of the wind respectively, at latitude $\phi$ and longitude $\theta$) while the second calculates the line of sight velocities caused by the planet's rotation ($R_p$ being the planet's radius at 1 bar, $z$ being an altitude at a given pressure level,
and $\Omega$ the planet's bulk rotation rate). We therefore self-consistently account for the effects of planetary winds and rotation on the high-resolution transmission spectra. The Doppler effects of orbital motion, which will equally impact each grid cell of the 3-D model, are accounted for in the data analysis process, described separately. 

\startlongtable
\begin{deluxetable*}{lcc}
\centering
\tablecaption{HD 189733 Modeled System Parameters \label{tab:planetstarpar}}
\tablehead{
\colhead{Parameter} & \colhead{Value} & \colhead{Units}
}
%stellar and planet model parameters
\startdata
Planet radius, $R_p$ & 8.693 $\times$ $10^7$ $\pm$ 1.72 $\times$ $10^6$ & m \\
Gravitational acceleration, $g$ & 19.5 & m s$^{-2}$ \\
Orbital rotation rate, $\omega_\mathrm{orb}$	&	$3.3 \times 10^{-5}$ & s$^{-1}$ \\
Irradiation temperature, $T_\mathrm{irr}$	& 1700	& K \\
Planet internal heat flux, $T_{int}$ & 100 & K \\
Optical absorption coefficient, $\kappa_{vis}$ & 4 $\times 10^{-3}$ & cm$^2$ g$^{-1}$ \\
Infrared absorption coefficient, $\kappa_{IR,0}$ & 1 $\times 10^{-2}$ & cm$^2$ g$^{-1}$ \\
Specific gas constant, $R$ & 3523 & J kg$^{-1}$ K$^{-1}$ \\
Ratio of gas constant to heat capacity, $R/c_p$ & 0.286 & ... \\
Stellar radius, $R_*$ & 5.600 $\times 10^8$ $\pm 1.11 \times 10^7$ & m \\
Star effective temperature, $T_{eff, *}$ & 5780 & K \\
Planet-star distance, $d$ & 2.16 $\times 10^9$ & m \\
Impact parameter, $b$ & 0.663 & ...
\enddata
\end{deluxetable*}

We calculate the transmission spectrum for a wavelength range of 2285 -- 2347 nm and a spectral resolving power $R = 2.5 \times 10^5$, which is then convolved to a lower resolution of $R = 1.0 \times 10^5$ for direct comparison to data described in Section \ref{sec:obs}. The dominant feature within this particular wavelength range (corresponding to the wavelength range of the CRIRES observations we will analyze using these models) is the strong, separated ``comb" of CO lines, with H$_2$O and CH$_4$ providing the next strongest sources of opacity \citep{Miller-RicciKempton2012CONSTRAININGTRANSIT}. We note that we updated our H$_2$O opacity line list in this work from the sources listed in Table 2 of \citet{lupu14} to HITEMP \citep{rothman2010}, which we found was necessary to generate a positive cross correlation signal for this molecule. 

In our previous studies \citep{Miller-RicciKempton2012CONSTRAININGTRANSIT,Rauscher2014THEJUPITERS}, we computed idealized models in which the stellar disk was assumed to have uniform intensity. However, in this work we are comparing our models to observations obtained at various times during transit, and it is therefore necessary to include the effects of stellar limb darkening. We employ the quadratic limb darkening model described in \citet{Brogi2016RotationSpectroscopy}:
\begin{equation} \label{eq:limb_darkening}
I(\mu) = 1 - u_1(1-\mu) - u_2(1-\mu)^2 
\end{equation}
where $\mu$ is the cosine of the angle between the line of sight and the normal to the stellar surface, and the limb darkening parameters $u_1$ and $u_2$ for HD 189733 are 0.077 and 0.311 respectively \citep{Claret2011Limb}. Using Equation~\ref{eq:limb_darkening}, we generate a pixelated model ``star" where each pixel has an intensity between 0 and 1. To generate transmission spectra from the GCM output, we determine which stellar pixel illuminates each grid cell of the planetary atmosphere at the orbital phase of a given observation. The incoming stellar intensity is then multiplied by the appropriate limb-darkened scaling factor, such that $I_{0} = I(\mu)$.

For each of the rotation models, at each point in the observed transit, we then calculate the observed flux by first subtracting the flux blocked by the planet and the atmosphere from the total stellar flux then adding back in the flux that passes through the atmosphere. The total integrated flux through the atmosphere is the sum of the intensities of each line-of-sight ray, multiplied by the 2-D projected solid angle of its associated atmospheric grid cell.

We have modeled our transmission spectra at 39 different consecutive times in transit that correspond to the observations described in Section \ref{sec:obs} and explored the different effects the planet's winds and rotation have on the spectra by turning on and off different physical aspects, creating wind-only, rotation-only, and combined-effects models. Additionally, we have modeled these spectra under two different chemical regimes. The first is a constant volume mixing ratio regime (VMR), in which molecular hydrogen, water, and carbon monoxide have constant abundances throughout the atmosphere at 99.8\%, 0.1\%, and 0.1\%, respectively. These values were selected to match the best fit values from the previous analysis of these data in \citet{Brogi2016RotationSpectroscopy}, and such fixed VMRs could be physically representative of a planet with quenched chemical abundances, in which atmospheric dynamics distribute molecular species faster than chemical reaction rates can return the gas mixture to thermochemical equilibrium (e.g. \citealt{Moses2011Chem, Venot2012HJAtmosphere, Venot2015Chem, Drummond2018HD189}). The second case considered is that of local chemical equilibrium (LCE), which is a reasonable first-order approximation for hot planets such as HD 189733b. 

\section{Model Results} \label{sec:modelres}

\subsection{GCM Results} \label{sec:gcmres}
\begin{figure}[!ht]
\centering
\includegraphics[width=0.5\textwidth]{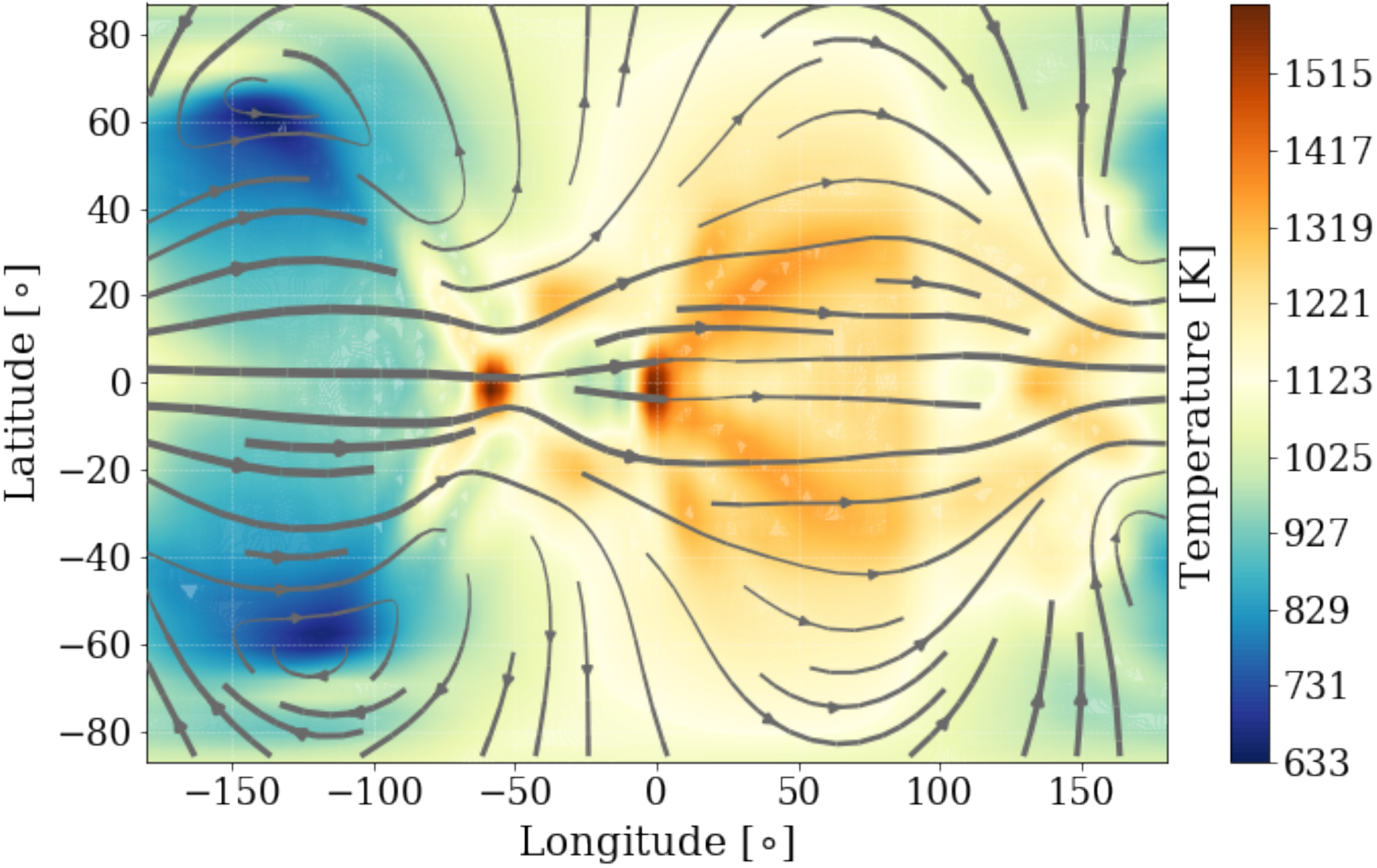}
\includegraphics[width=0.5\textwidth]{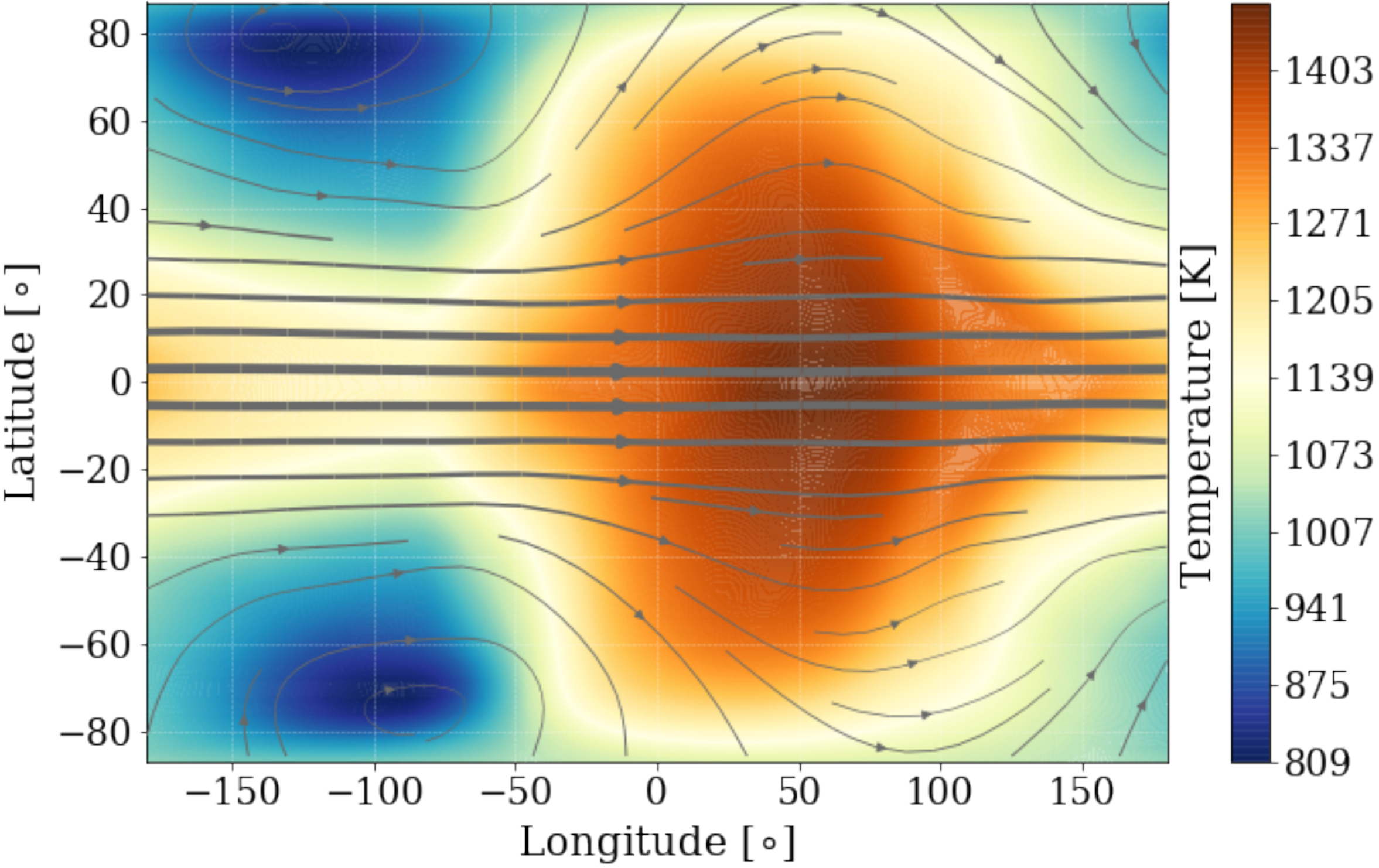}
\caption{Cylindrical projections of the temperature and wind fields (shown here as the stream function), from two different vertical levels in the planet's atmosphere, for the synchronous rotation case. The substellar point (zero latitude, zero longitude) is at the center of each plot. The thickness of the vectors is a function of wind speed. Top: the 0.1 mbar level, within the pressure range probed by transmission spectroscopy. The flow has two components: one that is from day to night, across the terminator, and one that is an eastward jet along the equator. The maximum wind speed at this level is 5.1 km/s. Bottom: the 130 mbar level, near the planet's infrared photosphere, showing the atmospheric structure influencing the planet's emission. One can clearly see the standard eastward, equatorial jet, which advects the hottest region of the atmosphere to be east of the substellar point. The maximum wind speed at this level is 4.3 km/s. Plots of the infrared photosphere and 0.1 mbar pressure level for all twelve models can be found in Appendix \ref{sec:appglobalmap}.}
\label{fig:IRP}
\end{figure}

The twelve models of HD~189733b we simulated, with the range of rotation rates shown in Table~\ref{tab:modelpar}, reproduce the circulation patterns we have come to expect from previous work. The model of HD~189733b with synchronous rotation shows the standard hot Jupiter circulation pattern \citep{Showman2011EquatorialExoplanets,Perez-Becker2013AtmosphericJupiters,Komacek2016AtmosphericDifferences}: the development of an eastward jet along the equator, which extends throughout most of the atmosphere. High in the atmosphere, for pressures less than $\sim$0.1 mbar, the eastward equatorial flow is accompanied by a significant direct substellar-to-antistellar flow. The top panel of Figure~\ref{fig:IRP} shows a map of the temperature and flow pattern in this regime, which is the pressure range preferentially probed by transmission spectroscopy. The bottom panel of Figure~\ref{fig:IRP} shows the temperature and wind pattern near the planet's infrared photosphere, where the equatorial jet is able to advect the hottest region of the planet eastward of the substellar point before it can cool efficiently. This is the atmospheric structure that influences observations of the planet in emission. The upper atmosphere and infrared photosphere maps for all models can be found in Appendix~\ref{sec:appglobalmap}.

\begin{figure*}[!ht]
\centering
\includegraphics[width=\textwidth, keepaspectratio]{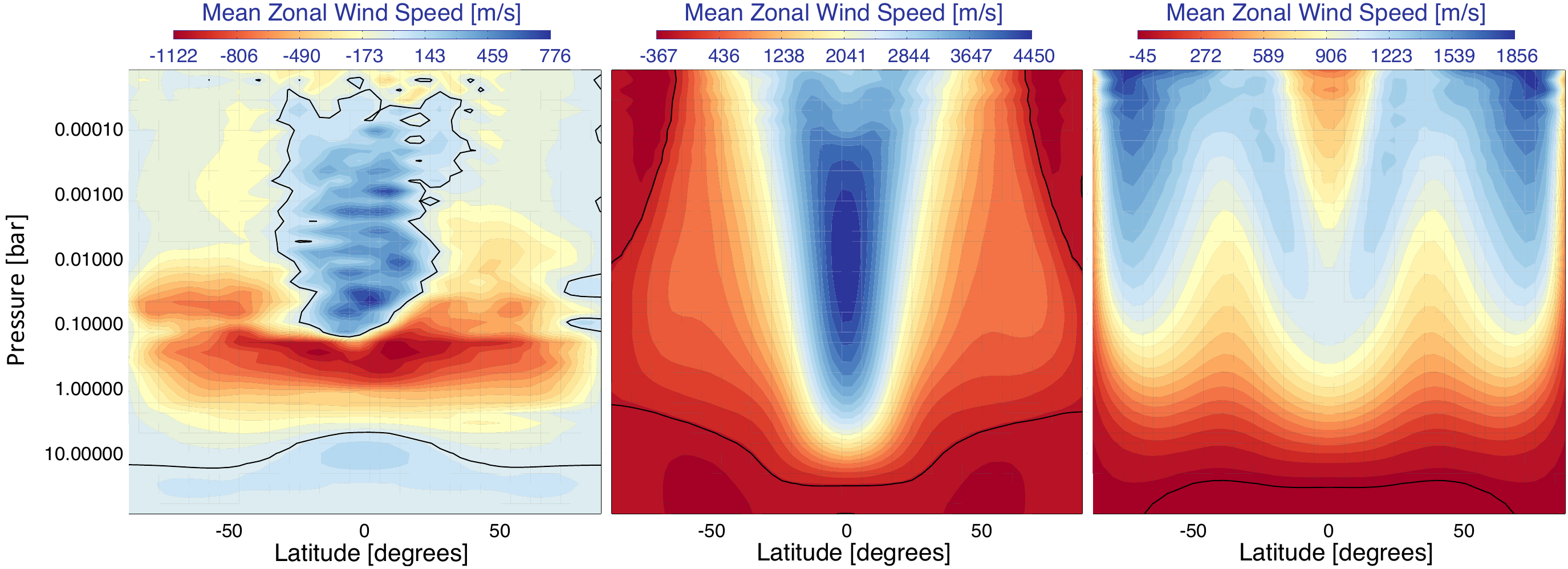}
\caption{The longitudinally averaged east-west winds, in the rotating frame of the planet (which differs between models), as a function of latitude and pressure throughout the atmosphere. Shown are the most slowly rotating of our models (left), the synchronously rotating case (middle), and the most quickly rotating model (right). The solid black line divides eastward (positive) from westward (negative) average wind speeds. In a more quickly rotating atmosphere there are multiple, thinner, eastward jets, in the synchronous case we find the standard hot Jupiter eastward equatorial jet, and in the slowest case the circulation pattern changes such that there is a significant westward component to the circulation. Plots for all twelve models can be found in Appendix \ref{sec:appzonalmap}.}
\label{fig:ZWM}
\end{figure*}

When we compare the synchronous case to models with faster and slower rotation rates, we also find the trends that we expect. The faster the rotation, the stronger the role of the Coriolis force in constraining the wind patterns, and so we see thinner jets and more of them, while in the more slowly rotating models (up through the $P_{rot}=4.69$ days model), the equatorial jet becomes wider in latitudinal extent. High in the atmosphere there is a more complex interaction between the jet pattern and the day-to-night flow, but the same general trends hold true. As the jets get narrower the wind speed of the jets also decreases (see Table~\ref{tab:modelpar}). This can simply be understood as horizontal shear limiting the latitudinal gradient in wind speed; for the same wind speed, the thinner jets would produce much stronger horizontal shear than the wider ones. These trends match those originally seen in the non-synchronous hot Jupiter models of \citet{Showman2009Atmospheric209458b} and \citet{Kataria2013Atmospheric}. An exception to this gentle trend in jet width and speed is seen for the two most slowly rotating models, where the circulation pattern is disrupted and becomes dominated by westward flow near the IR photosphere. This change in circulation regime for very slow rotators was originally discovered by \citet{Rauscher2014THEJUPITERS} and has been dynamically analyzed by \citet{PennVallis2017Thermal}. Figure~\ref{fig:ZWM} shows average profiles of the planet's east-west winds throughout the atmosphere, for the slowest, synchronous, and fastest rotation models; these plots for all twelve models can be found in Appendix~\ref{sec:appzonalmap}.

\begin{figure}[!ht]
\includegraphics[width=\linewidth, keepaspectratio]{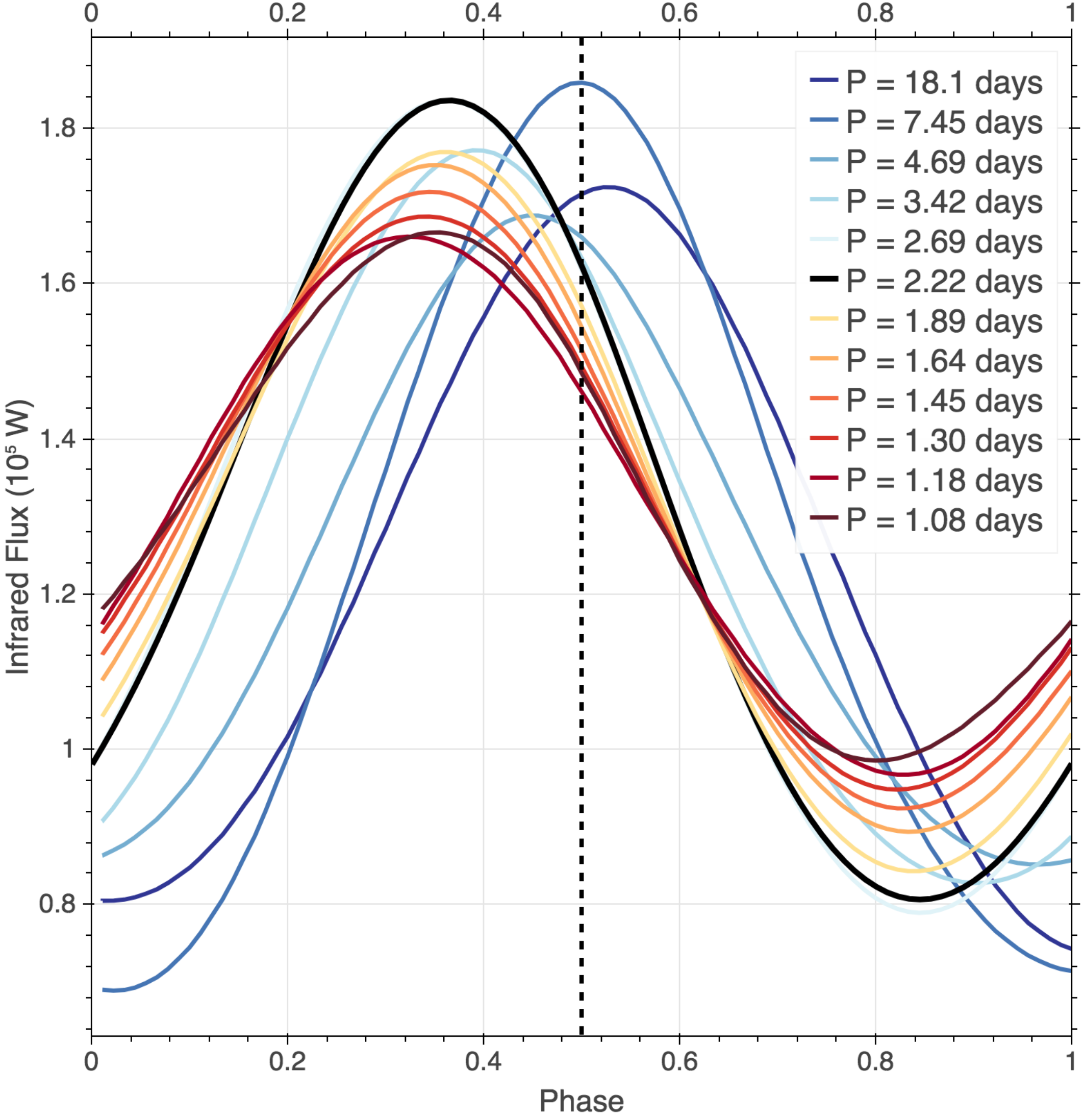}
\caption{Phase curves for all twelve GCM models, showing the predicted total infrared emission from the planet. Observed phase curves of HD~189733b exhibit a shift in peak to slightly before 0.5 (dashed black line). Thus, we can rule out the models with rotation periods longer than 4.69 days, where we observe a shift the opposite direction.}
\label{fig:phase}
\end{figure}

The disrupted circulation pattern of the two most slowly rotating models has an immediate observable implication: the westward flow forces the hottest region of the atmosphere to be at---or even westward---of the substellar point. In Figure \ref{fig:phase} we show simulated flux curves of planet emission as a function of orbital phase, and the slowest rotating models are notably different in when the peak flux occurs. The planet HD~189733b has been observed in thermal emission, at multiple wavelengths, and those data all show a peak in flux that occurs before secondary eclipse \citep{knutson12}, which corresponds to an orbital phase of 0.5 and is when the substellar point on the planet is facing directly toward the observer. For all of our models with rotation faster than $P_{rot} = 7.45$ days, the hot spot is shifted to the east of the sub-stellar point, resulting in phase curves that agree with observations, but the two most slowly rotating models are inconsistent with the data. Although phase curves are useful in ruling out these slower cases, the shapes of the remaining phase curves are subject to degeneracies \citep[e.g.,][]{Showman2009Atmospheric209458b}, making it difficult to constrain the rotation rate. Thus, high resolution spectroscopy that can detect the Doppler signatures of winds and rotation is a more direct means of determining the rotation rate of a planet.

While the bulk rotation of the planet will induce a broadening of transmission line profiles, the atmospheric dynamics will also contribute to the width and shift of lines, depending on the particular pattern of winds in the region of the atmosphere probed by these observations. The absorption lines we measure in this wavelength range should originate from pressures between roughly 0.01-1 mbar \citep{Miller-RicciKempton2012CONSTRAININGTRANSIT}. At these pressure levels the flow is a combination of a day-to-night component and the eastward equatorial jet, for the synchronously rotating model (see Figure~\ref{fig:IRP}). The upper atmosphere flow patterns for all models, shown in Appendix \ref{sec:appglobalmap}, demonstrate that the particular details of the wind structure are strongly influenced by the planet's rotation rate. As Table \ref{tab:modelpar} reports, the wind speeds can exceed the planet's bulk rotational speed, meaning that the complex velocity pattern due to the atmospheric circulation plays an integral role in shaping the observed spectra.

\begin{figure*}[!ht]
\centering
\includegraphics[width=\textwidth, keepaspectratio]{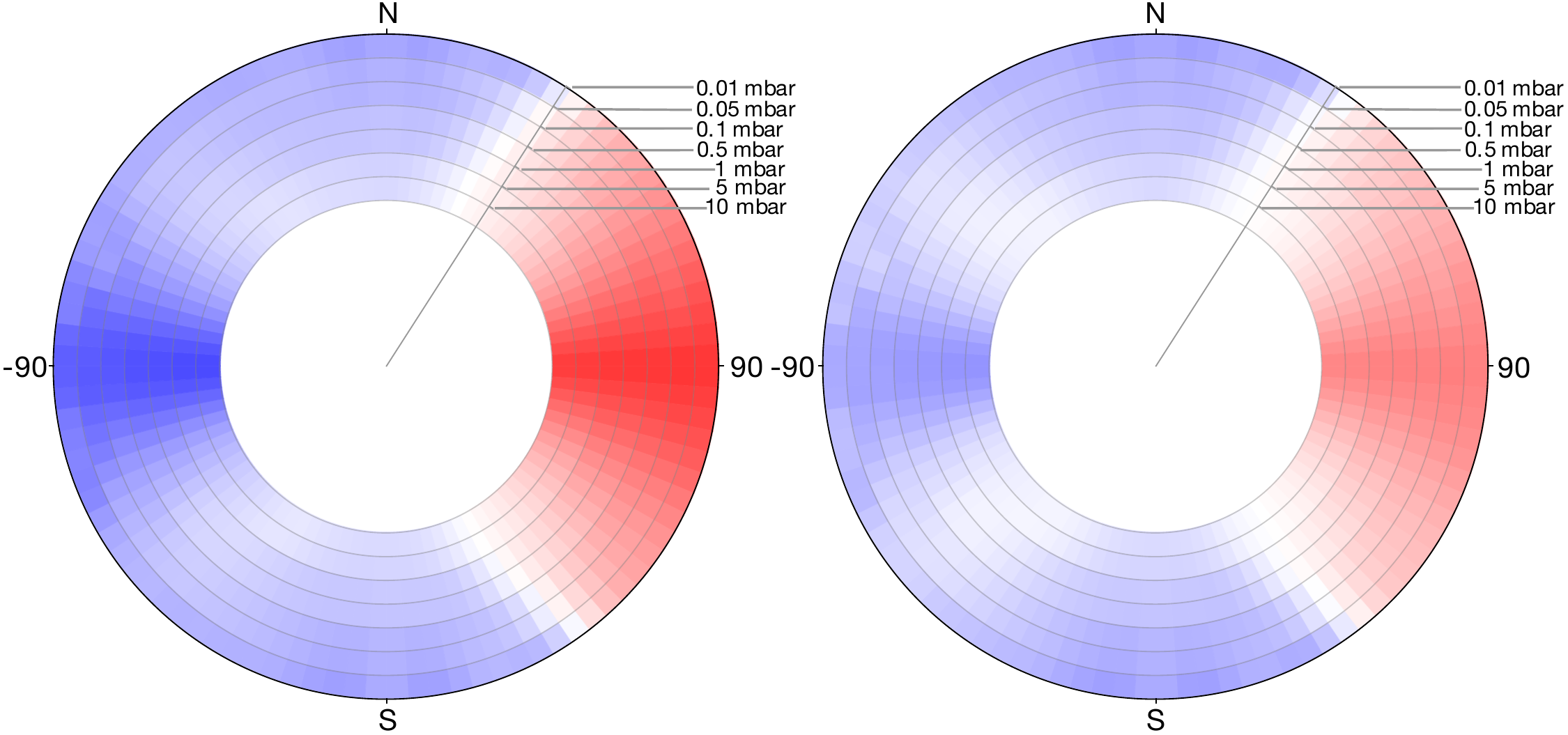}
\includegraphics[width=0.5\linewidth, keepaspectratio]{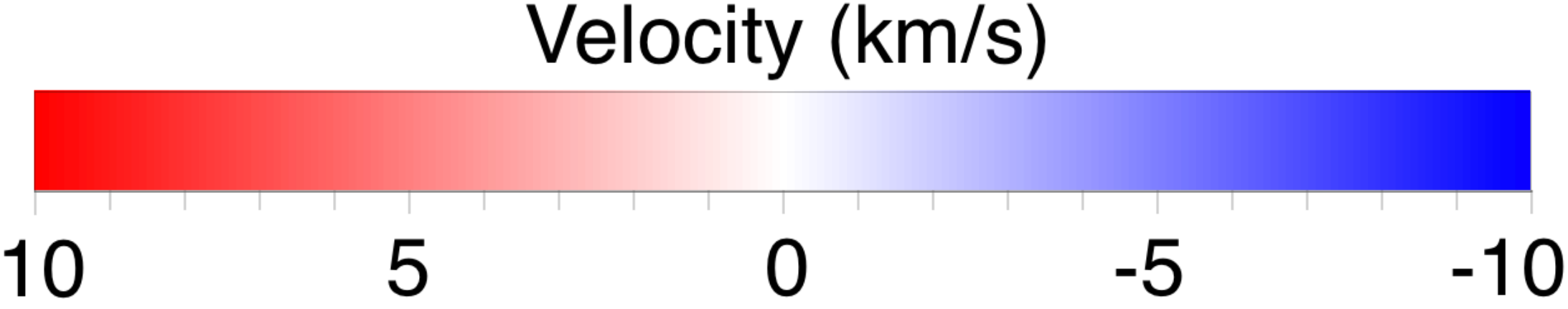}
\caption{Cross-section along the planet's terminator showing the line-of-sight velocities toward an observer during transit for the synchronously rotating model. Left: the full velocity field, from the combined effects of winds and rotation. Right: the line-of-sight velocity field due only the winds in the planet's atmosphere. For this model the eastward equatorial jet and the day-to-night flow over the poles have comparable speeds; the equatorial component is further enhanced by the rotation of the planet. Only the outermost layers of the atmosphere (those that are probed by transmission spectroscopy) are shown. Note: the radial extent of the atmosphere is not to scale. Velocity maps for all 12 of the models can be found in Appendix \ref{sec:applos}.}
\label{fig:LOS}
\end{figure*}

Figure~\ref{fig:LOS} helps to visualize the relative contributions of the wind and rotational velocity structures to the line-of-sight Doppler shifts probed in transmission. We show the local line-of-sight velocities, for a cross-section of the atmosphere through the planet's terminator, with the planet oriented as during transit (and with the radial scale of the upper atmosphere enlarged to show spatial detail). The path that a light ray takes in transmission through the planet's atmosphere crosses multiple longitude and altitude locations, so the full three-dimensional calculation is needed to fully capture the Doppler shifts in detail, but the terminator cross-section is a good first approximation. In particular, this shows that the high-altitude wind pattern in the synchronous model has equatorial eastward flow at similar speeds to the day-to-night flow over the poles; the former component is roughly doubled by the contribution of the rotational velocities, while the flow near the poles is minimally affected.

In Appendix~\ref{sec:applos} we show the line-of-sight velocity cross-sections for all twelve models, with the combined wind and rotational components. As indicated by Table~\ref{tab:modelpar}, we see from these patterns that the rotational velocities should dominate the Doppler signal in the most quickly rotating models, while the wind pattern is the main contributor for the more slowly rotating models. While the prevalent equatorial eastward jet contributes to Doppler broadening in an additive way with the rotational field, these plots show that the substellar-to-antistellar flow component should preferentially induce a blueshifted signal, especially for latitudes nearer to the poles. We may also expect that the spatial structure of the line-of-sight velocities may lead to anomalous Doppler shifts when the planet is at or near the edge of the limb-darkened star, as some parts of the atmosphere will be preferentially illuminated.

\subsection{Transmission Spectrum Results}\label{sec:transspecres}

\begin{figure*}[!ht]
\centering
\includegraphics[width=\textwidth]{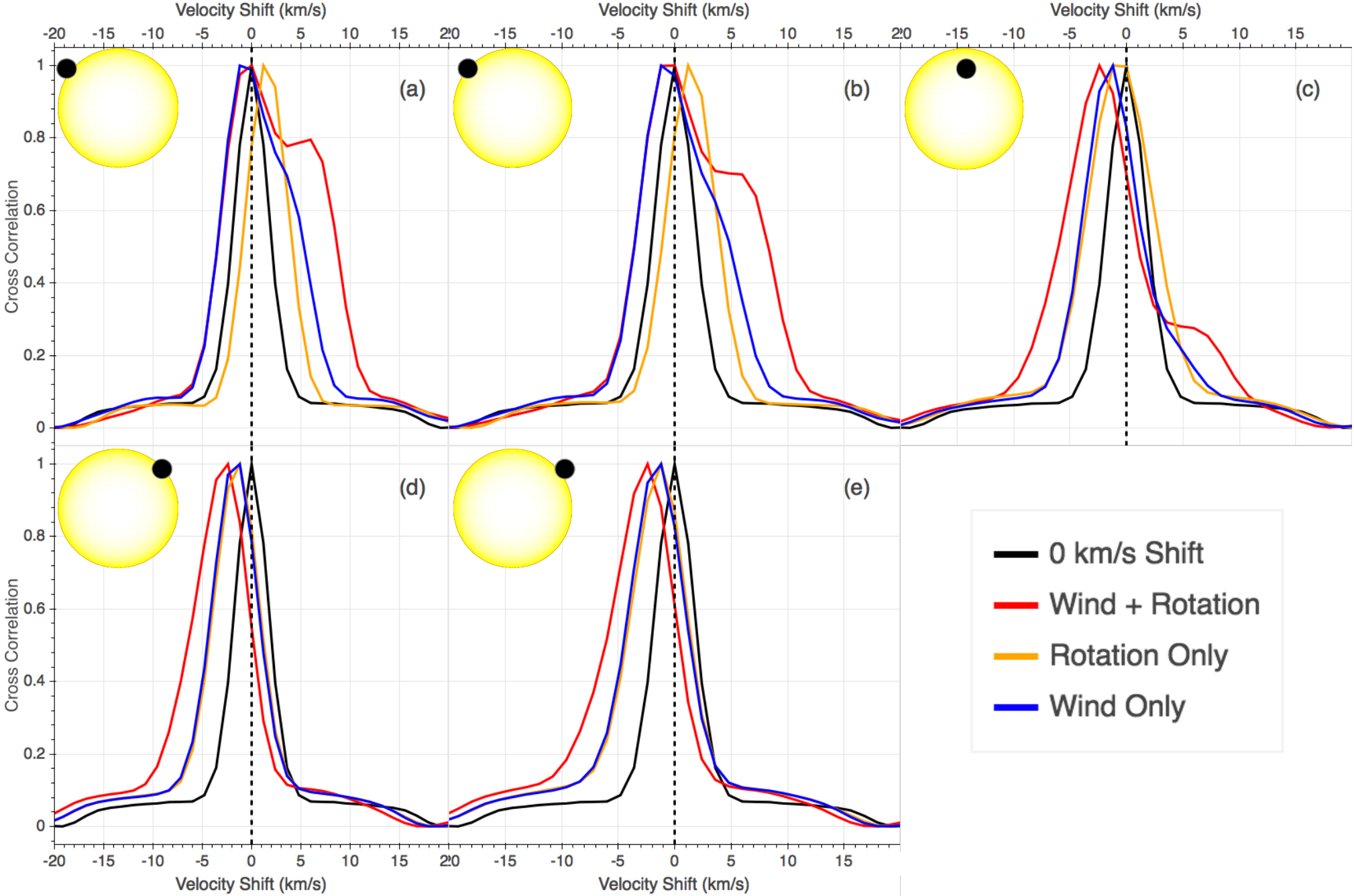}
\caption{Cross-correlation of LCE models with stationary case for, from left to right, early ingress (a), mid ingress (b), center transit (c), mid egress (d), and late egress (e) in the tidally synchronous model at the spectral resolution of CRIRES. The shape of the CCFs clearly changes depending on the point in transit, as different regions of the atmosphere are back-illuminated by the star. This is also apparent when looking at the CCFs for different rotation cases, as different models have different atmospheric architectures (see Appendix \ref{sec:appccf} for the center transit CCFs for each of the twelve models).}
\label{fig:CCF}
\end{figure*}

Prior to comparing our models directly to the observed data, we first perform ``model-on-model'' cross correlations of our model spectra against a narrow rest-frame template (no Doppler shifts applied from winds or rotation) to separately assess the effects of winds and rotation on each of our calculated spectra. The rest-frame template used for this entire analysis is a model produced at the center of transit (impact parameter of zero). Later, when comparing the models to data in Section~\ref{sec:ccf}, we will use the full 3-D consistent model spectra as the cross correlation templates. We perform the model-on-model analysis for five points in HD 189733b's observed orbit (early ingress, mid ingress, central transit, mid egress, and late egress), as shown in Figure~\ref{fig:CCF} for the tidally-synchronous case. The ingress and egress spectra differ from the center-transit spectra in that only a portion of the planet is in front of the host star, which can substantially bias the cross correlation signal relative to what would be expected during transit ``totality", when the planet is fully in front of its star. For example, during ingress the rotationally receding limb of the planet occults the star, which brings about an excess red-shift in the net cross correlation velocity.

The cross correlation process has greatest sensitivity to the strongest absorption features, which are optically thick at pressures $\lesssim 1$ mbar. As a result, the transmission spectrum primarily probes the velocity field in the upper atmosphere, which is dominated by day-to-night winds. This is apparent in the cross correlation functions, which tend to be strongly peaked at blue-shifted velocities for each of the center-of-transit models (Figure~\ref{fig:appccf}). Separately analyzing the contributions due to winds and the planet's rotation, we find net blue-shifts due to winds in each of the models at the $\sim 1 - 2$ km/s level, regardless of rotation rate. The equatorial jet also contributes to the Doppler shift signature of the winds. This provides a broadening component to the shape of the cross correlation function (CCF) resulting from the approaching and receding components of the jet across opposing limbs. The planet's rotation provides an additional broadening component to the CCF for cases where the rotational velocities are larger than the wind velocity dispersion. This occurs for all of the models with $P_{rot} > 2.6$ days. For the slowest rotating models, the velocity dispersion from winds exceeds that of the rotational motion, and the width of the CCFs is set by the winds alone. A key implication is that signatures of planetary rotation are not expected to be identifiable from cross correlation analysis alone if the planet is rotating sub-synchronously, because of the dominant effects of atmospheric winds. 

One might expect the effects of rotational broadening to be symmetric about a velocity shift of zero, but from Figure~\ref{fig:appccf}, we see that the faster rotating models produce CCFs that are both broader, but also more asymmetric toward blue-shifted velocities. This comes about for two reasons. First, the blue-shifted limb of the planet is also the hotter one (because the hot-spot leads ahead of the sub-stellar point), which causes the approaching limb to also be more inflated. In contrast, the red-shifted limb is cooler and less inflated, and therefore occults a geometrically smaller portion of the host star. Secondly, the cooler red-shifted limb also has a lower abundance of CO (and a correspondingly greater abundance of CH$_4$), which leads to a weaker cross correlation signal coming from that side of the planet. 

We note that these cross-correlation functions obtained from noiseless model spectra cannot be compared directly to CCFs obtained by cross correlating against observed data. The analysis that we apply to observations also produces distortions in the planet spectrum, which could potentially mimic signatures of atmospheric circulation or be partially confused with them. Consequently, in the next Section we explain how we process real observations to correct for these effects of our data analysis technique.

\section{Observations} \label{sec:obs}
The data utilized in this study are described in detail in \citet{Brogi2016RotationSpectroscopy}. They consists of a sequence of 45 high-resolution, near-infrared spectra observed at the CRIRES spectrograph \citep{Kaeufl2004CRIRES:VLT} on July 30, 2012, as part of DDT program 289.C-5030. The spectrograph, now decommissioned, was mounted at the ESO Very Large Telescope UT1 in Chile and operated at a resolution of $R\sim 100,000$ with a 0$^{\prime\prime}.2$-wide slit, covering the range 2287.5 – 2345.4 nm in the configuration adopted here.

The observations were carried out by nodding along the slit via an ABBA pattern for accurate subtraction of thermal background and emission lines in the Earth's atmosphere. With an exposure time of 60s per nodding position, we obtained 90 spectra, or 45 couples of AB or BA pairs. These are combined through the standard CRIRES pipeline and the corresponding 45 one-dimensional spectra extracted. We obtain an average signal-to-noise ratio (S/N) of 225 per pixel (1 pixel corresponds to approximately 1.5 km s$^{-1}$ in velocity space) in spectral regions free from absorption lines.

The subsequent calibration of the data is described in \citet{Brogi2016RotationSpectroscopy} and consists in realigning all the spectra to a common wavelength scale, which is chosen to be that of the absorption lines in the Earth's atmosphere (telluric absorption). In this way, telluric lines always fall on the same spectral channel, and they also provide an accurate wavelength calibration for our spectra, with typical precision of 50-75 m/s per line. 

We depart from \cite{Brogi2016RotationSpectroscopy} for the correction of the stellar spectrum. Stellar CO lines are distorted by the transit of the planet producing the well-known Rossiter-McLaughlin (R-M) effect. Although the effect on the centroid of stellar lines is on the order of tens of m/s, once the average line profile is removed by our analysis the actual radial-velocity change in the residual stellar component spans $\pm v\sin(i_\star)$ during transit, which is $\pm 3.3$ km s$^{-1}$ in the case of HD 189733. The amplitude of these residuals is 0.5-1.0 percent, which is sufficient to outshine the atmospheric signature of HD 189733b. We therefore proceed to model the R-M effect as in \citet{Brogi2016RotationSpectroscopy}, but using as input a spatially-resolved, 3-D stellar model matching the properties of the star (5000 K, $\log(g)$ = 4.5, [Fe]/[H] = $-0.03$). 
This was produced using the state-of-the-art realistic three-dimensional radiative hydrodynamical (RHD) simulations of stellar convection carried out with the Stagger-code
\citep{Nordlund2009SolarConvection}. The RHD simulations grid cover a large part of the HR diagram, including the evolutionary phases from the main sequence over the turnoff up to the red-giant branch for low-mass stars \citep{Magic2013TheModels}. The simulation used in this work
has been employed to compute synthetic spectra in the same spectral range as the observations with the multidimensional pure-LTE radiative transfer code \texttt{Optim3D} \citep{Chiavassa2009RadiativeObservations,Chiavassa2018STAGGER}. The code takes into account the Doppler shifts occurring due to convective motions and solves monochromatically for the emerging intensity, including extensive atomic and molecular continuum and line opacity data from UV to far-IR. 

At the resolution of CRIRES, it is possible to isolate a set of stellar CO lines far from telluric absorption and visually verify that the chosen stellar parameters are appropriate, i.e., the modeled spectral lines have the correct depth and shape. Using a model for the stellar spectrum is a major improvement from our previous study, which instead relied on parameterizing an average stellar line profile through micro- and macro-turbulence, and was inadequate to reproduce the complicated velocity fields in the convective envelope of the star. 

We could not properly correct the spectra falling on detector 3 of CRIRES with the above analysis. Given the high density of both stellar and telluric lines in that spectral region, it is likely that our algorithm to reconstruct the instrument profile through Singular Value Decomposition fails, leading to an imperfect correction of stellar lines. Since detector 4 is also excluded due to a well documented problem with the differential gain of odd and even columns, we are left with only half of the spectral range of CRIRES for this study (detectors 1 and 2, spanning 2287.5-2316.0 nm).

After dividing out the model stellar spectrum, telluric lines were removed and the residual spectra cross-correlated as in \citet{Brogi2016RotationSpectroscopy} to combine the signal of all the individual absorption lines in the planet transmission spectrum into one cross-correlation function (CCF).

\section{Data Cross Correlation} \label{sec:ccf}

Before comparing the GCM models to these data, we repeat the analysis of \citet{Brogi2016RotationSpectroscopy}. This is necessary to verify consistency, since our procedure to correct for the stellar spectrum has changed. The only other difference with the previous analysis is that we increase the resolution in the semi-amplitude of the planet's orbital radial velocity ($K_\mathrm{P}$), and we use a new set of values for the planet rotational equatorial velocity ($V_\mathrm{eq}$) consistent with the input parameters for this work (Table~\ref{tab:modelpar}, right column). In this analysis, the data are always cross-correlated with a {\sl narrow} template for the planet transmission spectrum, which is a model not including the velocity field due to winds and rotation (this is the center-transit model already described in Section \ref{sec:transspecres} and utilized to obtain the theoretical CCF of our models). Since the cross-correlation operator is similar to a convolution (i.e., it broadens the planet line profile), this choice ensures that we apply a consistent level of broadening throughout the analysis. In order to evaluate the goodness of fit of models with varying circulation patterns, we need to estimate how the true planet line profile is broadened by cross correlation and possibly altered by our data analysis. Following \citet{Brogi2016RotationSpectroscopy}, we inject each model in the data immediately after wavelength calibration and pass the injected signals through the pipeline. To ensure that the injection does not sensibly alter the analysis, models are scaled down to line contrast ratios of at most $10^{-4}$ compared to the stellar continuum. The CCF of the injected signal (CCF$_\mathrm{inj}$) will contain the model, the real signal, and noise. The CCF of unaltered data (CCF$_\mathrm{real}$) will instead only contain real signal and noise. Since the noise remains constant due to the very small injection, we can apply the linearity of the cross-correlation operator and obtain the CCF of the injected model via
\begin{equation}
\mathrm{CCF}_\mathrm{mod} = \mathrm{CCF}_\mathrm{inj} - \mathrm{CCF}_\mathrm{real},
\end{equation}
CCF$_\mathrm{mod}$ incorporates the same broadening and distortions of the real signal. It can thus be compared to the CCF of the real signal through chi-square analysis.
As outlined in detail in \citet{Brogi2016RotationSpectroscopy}, we assign a significance to each model based on how much more favorable its CCF is compared to a straight line. The latter indicates zero signal by definition, which occurs when the sample of cross-correlation values is consistent with a Gaussian distribution.

\begin{figure*}[!ht]
\includegraphics[width=\textwidth, keepaspectratio]{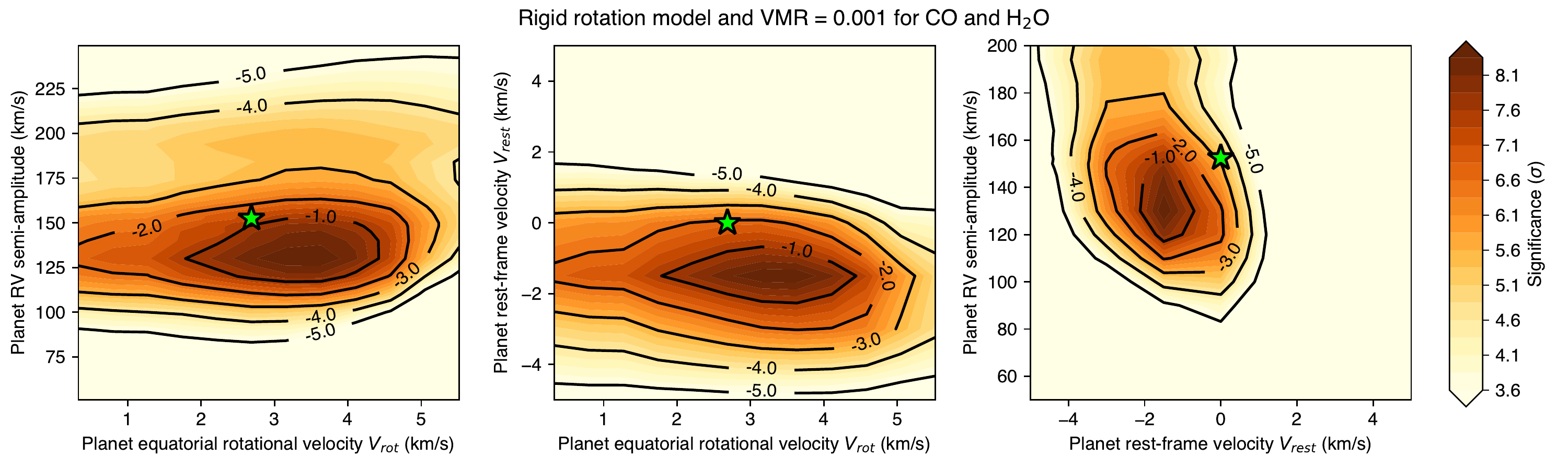}
\caption{Analysis of \citet{Brogi2016RotationSpectroscopy} repeated with our new three-dimensional radiative hydrodynamical simulation of stellar convection for the parent star and a new algorithm for removing stellar lines. As explained in Section~\ref{sec:ccf}, we obtain a more significant detection and tighter confidence intervals. The three panels show two-dimensional significance maps of the explored planet parameters: orbital radial-velocity semi-amplitude ($K_\mathrm{P}$), equatorial rotational velocity ($V_\mathrm{rot}$), and rest-frame velocity ($V_\mathrm{rest}$). The latter should be zero if all the velocity components have been accounted for. We measure instead a significant residual blue shift, discussed in Section~\ref{sec:ccf}. The green star denotes the literature value of the RV semi-amplitude (with error bars negligible compared to the scale of the plots) an assume a tidally locked planet with no excess rest-frame motion due to winds or other effects.}
\label{fig:remco_rigid}
\end{figure*}

Figure~\ref{fig:remco_rigid} shows the confidence intervals for $K_\mathrm{P}$, $V_\mathrm{eq}$, and for the planet rest-frame velocity ($V_\mathrm{rest}$). Despite using only 2/3 of the data of \citet{Brogi2016RotationSpectroscopy}, we obtain a stronger detection of the combined absorption of CO and H$_2$O at a significance of 8.6$\sigma$. This indicates that by improving the correction of the stellar spectrum we are also preserving a larger fraction of the planet signal, especially the CO spectrum. Qualitatively we confirm that our measurements are compatible with synchronous rotation, but the stronger detection and better cleaning of stellar residuals allow us to improve the confidence intervals in all three parameters. We measure $V_\mathrm{eq} = 3.4^{+1.0}_{-1.5}$ km s$^{-1}$ (previously $V_\mathrm{eq} = 3.4^{+1.3}_{-2.1}$ km s$^{-1}$), and rotation is now favored over models without rotation at 2$\sigma$ (previously 1.5$\sigma$). Rotational periods shorter than 1.2 days are now excluded at $>3\sigma$. The signal peaks at a planet rest-frame velocity of $V_\mathrm{rest} = -1.4^{+0.8}_{-0.9}$ km s$^{-1}$ (previously $-1.7^{+1.1}_{-1.2}$ km s$^{-1}$), which corresponds to a global blue shift in the planet cross-correlation function detected at 1.9$\sigma$. This is expected in the presence of day-to-night side winds since these are not included in the rigid-rotation model. Finally, we measure a planet radial velocity semi-amplitude of $K_\mathrm{P} = 131^{+22}_{-14}$ km s$^{-1}$, compared to the previous value of $K_\mathrm{P} = 194^{+19}_{-41}$ km s$^{-1}$, and still consistent with the literature value of $K_\mathrm{P} = 152.3^{+1.3}_{-1.8}$ km s$^{-1}$ computed in \citet{Brogi2016RotationSpectroscopy}.

\begin{figure*}[!ht]
\includegraphics[width=\textwidth]{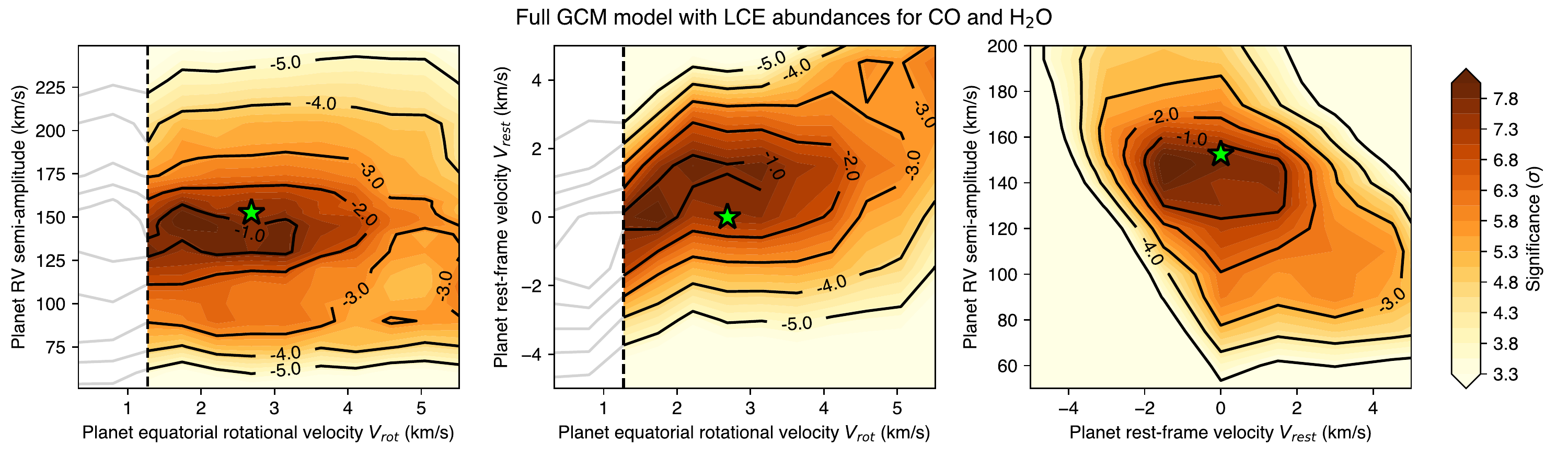}
\includegraphics[width=\textwidth]{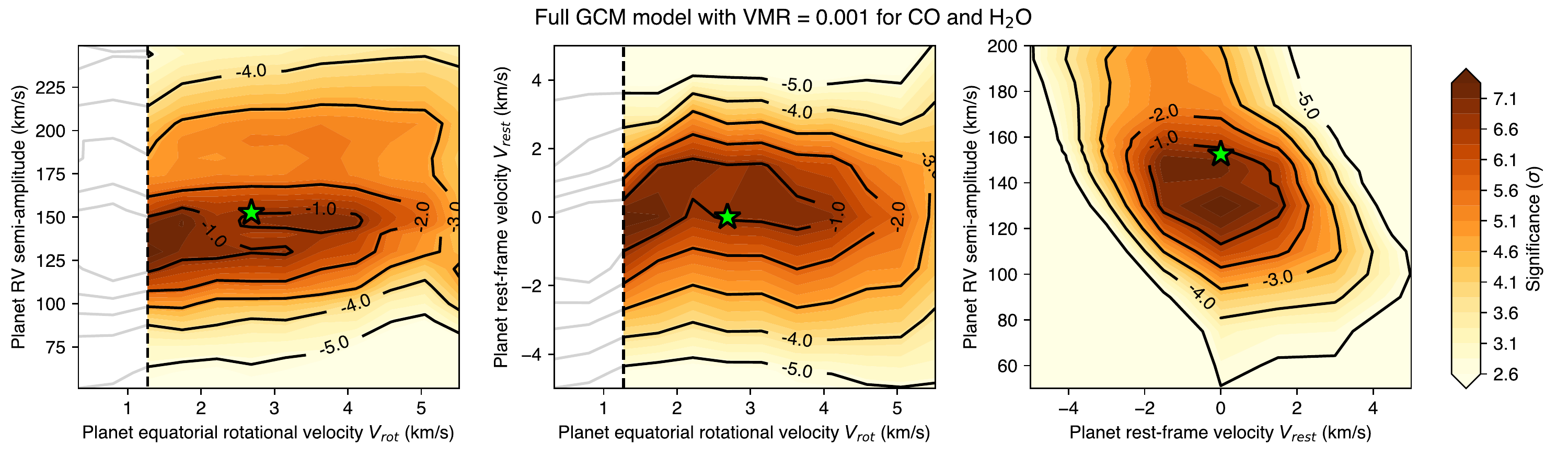}
\caption{Significance maps analogous to Figure~\ref{fig:remco_rigid}, but now obtained with the full GCM models described in this paper. The top panels are obtained with water and CO abundances in local chemical equilibrium (LCE), whereas in the bottom panels the volume mixing ratios are forced to be 10$^{-3}$, i.e. the best fitting value of \citet{Brogi2016RotationSpectroscopy}. The blanked-out region (left of the dashed vertical line) is excluded because our GCM models fail to reproduce the observed phase offset of the thermal emission peak for such slow rotations.}
\label{fig:gcm_full}
\end{figure*}

We now proceed to compare the models computed in this paper with the data (see Figure~\ref{fig:gcm_full}). We obtain a detection at a 8.2-$\sigma$ confidence level for the models in LCE (top panels), whereas the models with VMR of 10$^{-3}$ for CO and H$_2$O deliver a lower detection at 7.4$\sigma$ (lower panels). The middle and right panels clearly show that the planet signal retrieved with the full GCM models now peaks at zero rest-frame velocity. This is a strong indication that our 3D models correctly predict the global atmospheric circulation at the terminator, in particular the flow at relatively high altitude that produces a net blue shift. In contrast, from the left and middle panels, we note that the rotational rate of the planet becomes poorly constrained, except perhaps for very fast rotations. 

\begin{figure*}[!ht]
\includegraphics[width=\textwidth]{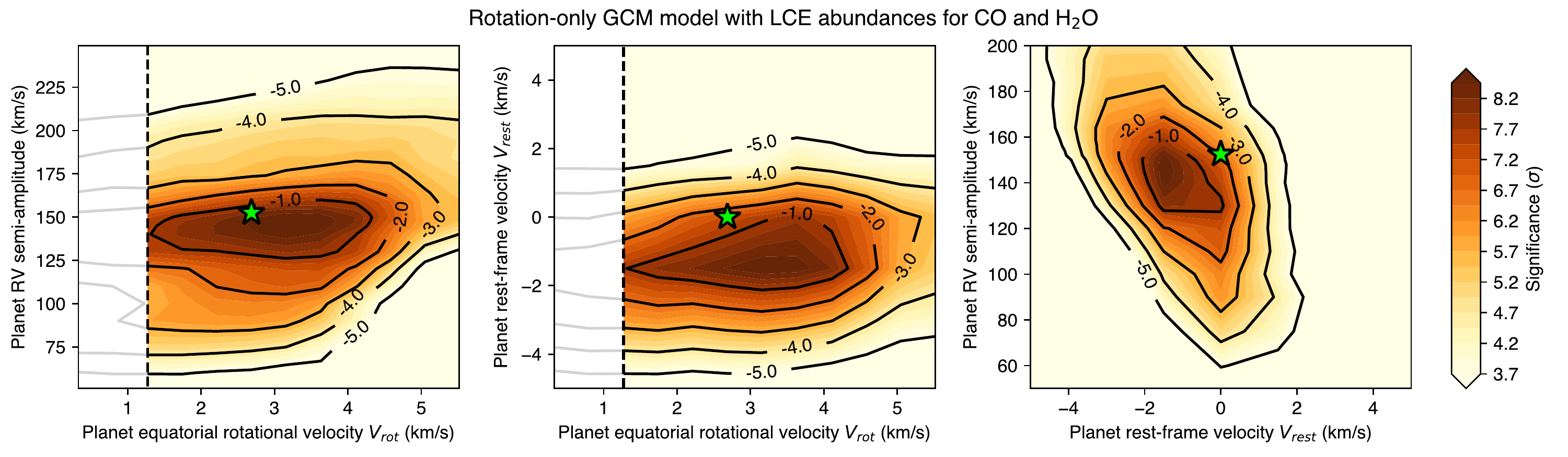}
\includegraphics[width=\textwidth]{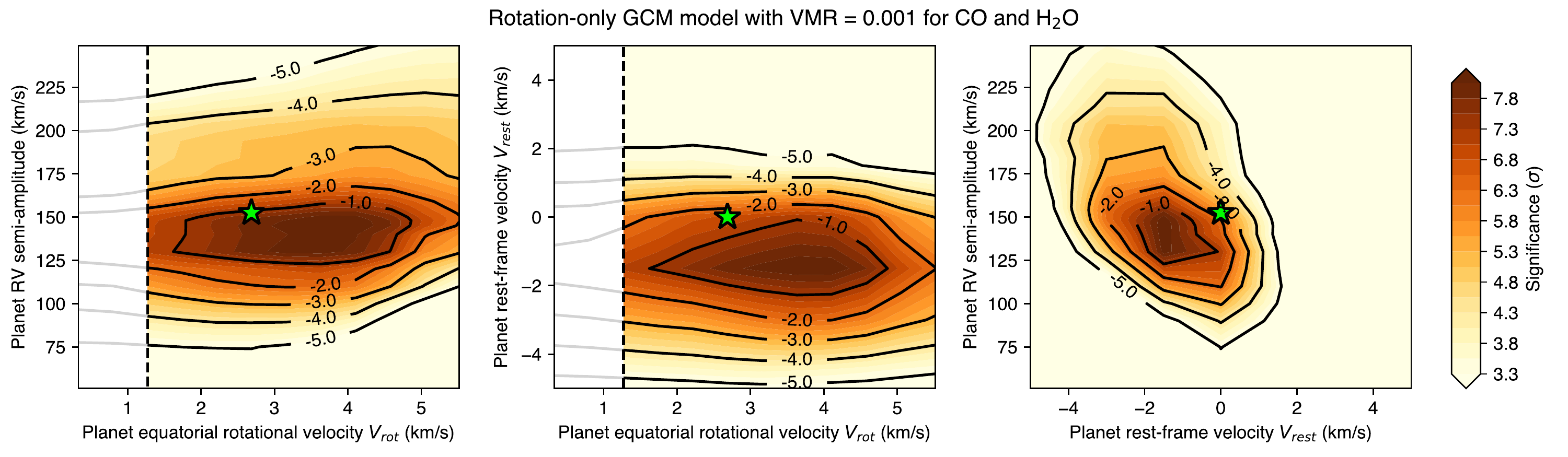}
\caption{Significance maps analogous to Figure~\ref{fig:gcm_full}, but including only the effects of rotation and neglecting additional circulation patterns. Confidence intervals are qualitatively similar to those obtained with the rigid-rotation model (shown in Figure~\ref{fig:remco_rigid}) including the ability of excluding slow and fast rotational periods and the residual measured blue-shift.}
\label{fig:gcm_rotonly}
\end{figure*}

Lastly, in Figure~\ref{fig:gcm_rotonly} we show the results from the same GCM models when the atmospheric circulation is turned off. These models hence only contain the effects of rotation. With these models we would expect the same qualitative results as for the repeated analysis of \citet{Brogi2016RotationSpectroscopy}, except that the planet is not a rigid body anymore. Not only is the significance of the detection enhanced with this last set of models (8.7$\sigma$ and 8.3$\sigma$ for LCE and VMR, respectively), but we also recover the sensitivity to the synchronous rotational period and exclude both fast and slow rotations. Having suppressed winds in the planet's atmosphere, the signal is again retrieved blue-shifted, as expected when the global high-altitude winds are not modeled. 

\section{Discussion and future prospects}\label{sec:dis}
We have shown that models using fully integrated physics and computed from first principles deliver cross correlation signals as significant as those obtained with simple parametrized models, optimized through a chi-square grid search. This key result suggests that the main assumptions of these GCM models are sound, and that the global atmospheric flow is correctly reproduced. 

In particular for HD 189733b, we confirm the detection of an overall blue shift of the planet's transmission spectrum by $-1.4^{+0.8}_{-0.9}$ km s$^{-1}$. More importantly, since the planet signal is retrieved at zero rest-frame velocity when cross correlating with our GCM models, we are finally able to directly link this shift to the presence of high-altitude winds flowing from the day to the night side of the planet, confirming one of the main theoretical predictions about the main atmospheric flow in irradiated giant exoplanets. 

Furthermore, the onset of strong equatorial super-rotation significantly broadens the planet line profile even for very slow rotation (low equatorial velocity $V_\mathrm{rot}$). Consequently, it is challenging to constrain these slow rotational regimes, at least at the current spectral resolution and signal-to-noise ratio of the observations. This demonstrates the utility, effectiveness, and need for the use of model predictions in analyzing high-resolution transmission spectra of hot Jupiters. Models neglecting the interplay between atmospheric winds and the planet's rotation -- for instance by independently retrieving rotational period and equatorial velocity -- would measure much tighter but incorrect constraints on the planet's rotational rate, because they would not account for the fact that these two parameters are physically correlated. The slowest rotational rates can, however, be constrained by adding the information from photometric phase curves. We showed that the thermal emission peak in models with the longest orbital periods occurs after secondary eclipse, which is inconsistent with a pre-secondary eclipse peaks observed for not only HD 189733b, but a significant fraction of the hot Jupiters. 

Conversely, we confirm the ability to constrain fast rotational rates, which again dominate the broadening of the planet line profile regardless of the additional equatorial flow. This means that in the future it will be possible to expand these observations to systems of different age and orbital separations, and test under which regime(s) synchronous rotation breaks down. Non-tidally-locked giant planets, if they rotate as fast as the solar system giants, will be easily discriminated with equatorial velocities in excess of 10 km s$^{-1}$. These measurements will be key to constrain how interiors of giant planets react to tides, which is still poorly known for solar system planets.

Our analysis is currently limited to two possible chemical regimes, namely Local Chemical Equilibrium (LCE) abundances for solar metallicity or a slight super-abundance of water and carbon monoxide at VMR = 10$^{-3}$. We see some indication that LCE models match the data more closely, with an increase in detection significance of 0.8$\sigma$. However, we are not yet at a stage where we can explore the opacity effects of different molecular species and abundances, nor did we attempt to adjust tunable parameters in the GCM simulation, such as the numerical dissipation necessary to capture sub-grid sources of physical dissipation. \citet{Thrastarson2011RelaxationSimulations} discuss some of the nuances and uncertainties associated with numerical dissipation. Since the radiative timescales vary by orders of magnitude within the atmosphere, it is possible that the lowest pressure (highest altitude) regions are under-damped in comparison to the rest of the atmosphere.
Exploring large grids of the above parameters is still prohibitive due to the fact that these GCMs are computationally intensive.

In addition to physically-based models for planet atmospheric circulation, in this work we have also utilized physically based, three-dimensional models of the stellar spectrum. These can correctly reproduce variations in the stellar convective blue-shift and optical path along the disk, allowing us to model in unprecedented detail center-to-limb variations and the Rossiter-McLaughlin effect caused by the planet during transit. Correcting our spectra with such models has led to an increase of 1$\sigma$ in the significance of the detected planet signal, even though we only utilized 2/3 of the data of \citet{Brogi2016RotationSpectroscopy}. This highlights the importance of accurate and precise stellar models at high spectral resolution to maximize on the scientific return of high resolution observations of the exoplanets that orbit these stars. To that end, synergistic work between the stellar and exoplanetary communities is warranted.

In conclusion, we have shown that high resolution spectroscopy uniquely constrains both atmospheric circulation patterns and rotation rates. In addition to the constraints coming from broad-band photometry and spectro-photometry, we now have the ability of directly measuring wind speeds of a few km/s reflected in the shape of cross-correlation functions. 
With new and improved high resolution spectrographs coming online in the near future and more immediate follow-up high-resolution spectroscopic studies on new survey targets (such as bright planets discovered with NASA's Transiting Exoplanet Survey Satellite), it will be possible to implement analysis techniques such as the ones presented in this paper to characterize a homogeneous set of transiting planets around bright stars. With enough signal-to-noise and time resolution, we could potentially trace the minute differences in the terminator flow during the varying phases of a transit \citep{Miller-RicciKempton2012CONSTRAININGTRANSIT}.
Even for non-transiting planets, winds and rotation are expected to affect line shapes in emission spectroscopy \citep{Zhang2017Emission}, and we aim at testing these predictions in future studies. Ultimately, the advanced characterization methods that are currently only feasible for hot Jupiters will be employable on smaller and cooler planets, with the goal of characterizing potentially habitable worlds.

\acknowledgements
This work was supported in part by the National Science Foundation Award No.\ 1559988 and by NASA Astrophysics Theory Program grant NNX17AG25G. M.B. acknowledges partial support by NASA, through Hubble Fellowship grant HST-HF2-51336 awarded by the Space Telescope Science Institute.

%%Appendix - additional figures
\appendix \label{app}

In the following sections we include plots showing the results from each of our 12 differently rotating models. Section \ref{sec:appglobalmap} contains maps of atmospheric temperatures and winds, for the 0.1 mbar and infrared photosphere levels, equivalent to the one shown for the synchronous case in Figure \ref{fig:IRP}. Section \ref{sec:appzonalmap} shows the zonal wind patterns for all models, equivalent to Figure \ref{fig:ZWM} in the main text. We show the line-of-sight velocity cross-sections (as in Figure \ref{fig:LOS}) for all models in Section \ref{sec:applos}. Finally, Section \ref{sec:appccf} contains a single modeled center-of-transit cross-correlation function for each model, described in Section \ref{sec:transspecres} and shown in Figure \ref{fig:CCF} at several points during transit for the synchronous case.

\section{Maps of the temperature and wind structure in the upper atmosphere and at the infrared photosphere} \label{sec:appglobalmap}

\begin{figure}[H]
  \centering
  \includegraphics[width=\textwidth, keepaspectratio]{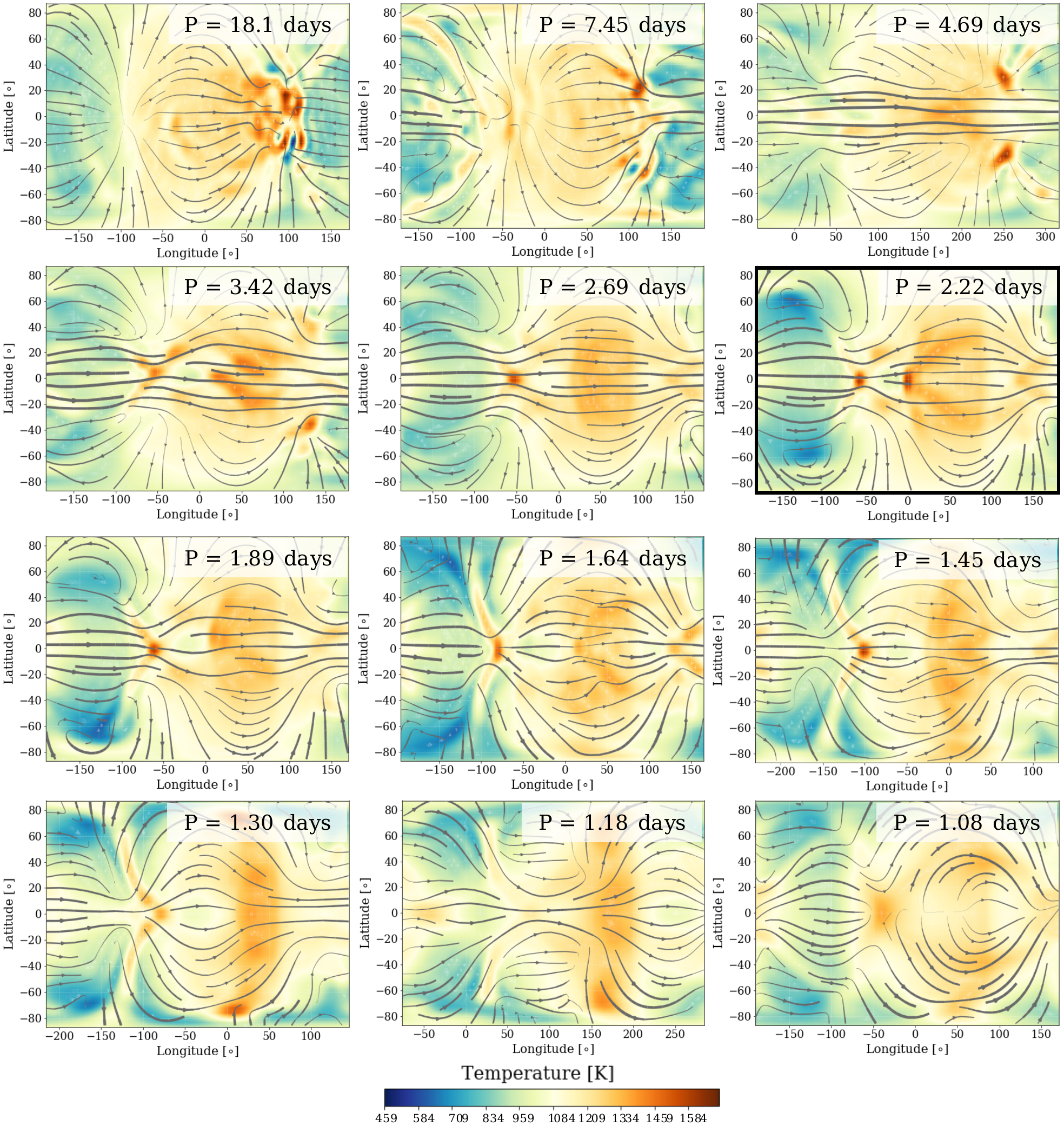}
  \caption{Global temperature and wind maps of the upper atmosphere (0.1 mbar) for all twelve rotation models. The maps have been shifted such that the substellar point (during transit for the non-synchronously rotating models) is at the center of the plot. The tidally synchronous case is outlined in black. At this pressure level and those higher in the atmosphere, the flow is composed of a direct substellar-to-antistellar component and the equatorial jet. At the fastest rotation rates the day-to-night flow competes with multiple eastward jets, while at the slowest rotation rates the eastward equatorial jet becomes disrupted and there is even strong westward flow across the night side.}
  \label{fig:my_label}
\end{figure}

\begin{figure}[H]
\centering
\includegraphics[width=\textwidth, keepaspectratio]{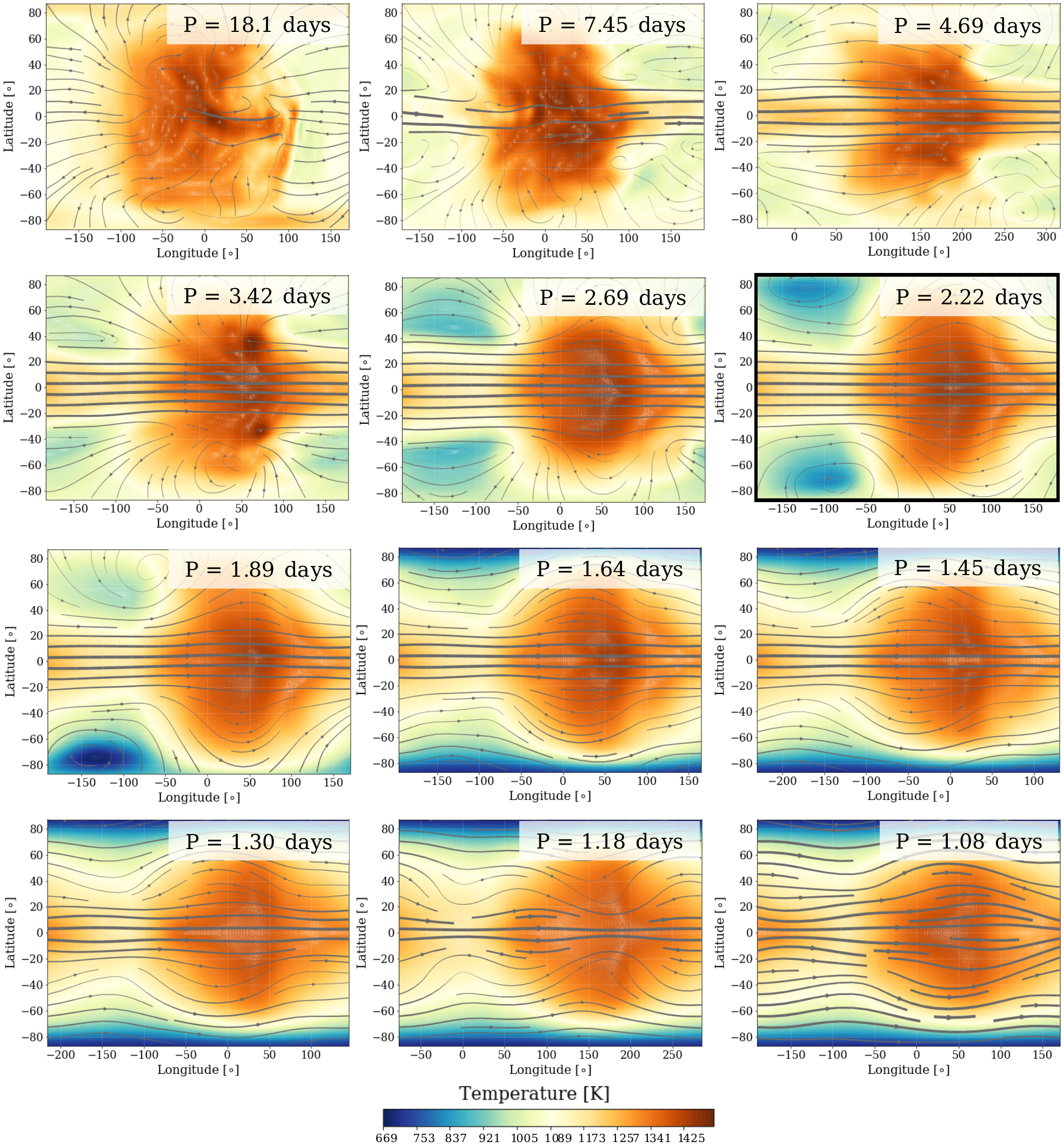}
\caption{Global temperature and wind maps of the IR photosphere (130 mbar) for all twelve rotation models. The maps have been shifted such that the substellar point (during transit for the non-synchronously rotating models) is at the center of the plot. The tidally synchronous case is outlined in black. Changing the rotation rate of the planet influences the pattern of winds, disrupting the standard eastward wind direction in the slowest cases. The day-night and equator-to-pole temperature gradients also differ between the different rotation models.}
\label{fig:appglobal}
\end{figure}
\section{Zonal wind maps} \label{sec:appzonalmap}
\begin{figure}[H]
\centering
\includegraphics[width=0.9\textwidth, keepaspectratio]{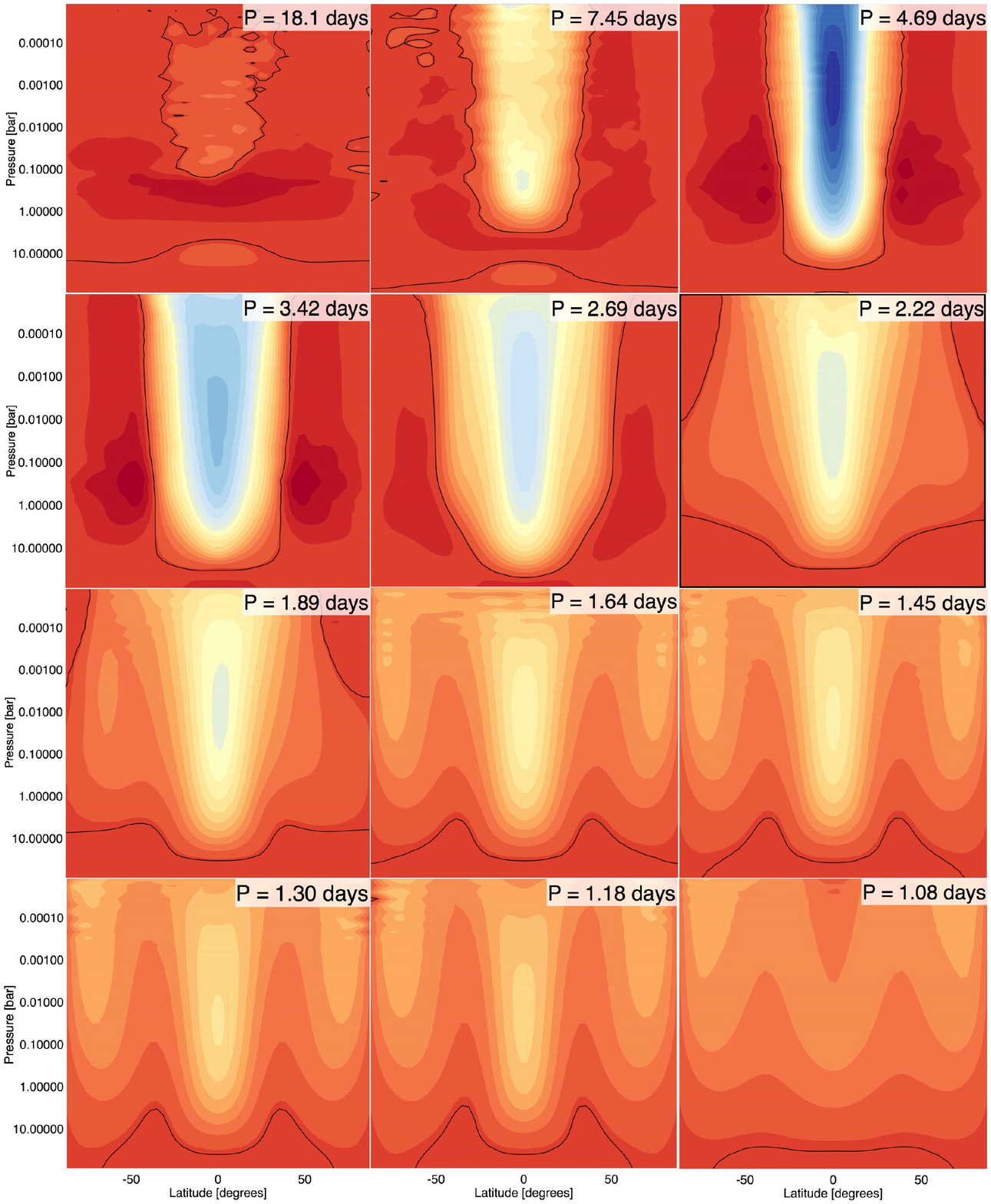}
\includegraphics[width=0.3\textwidth, keepaspectratio]{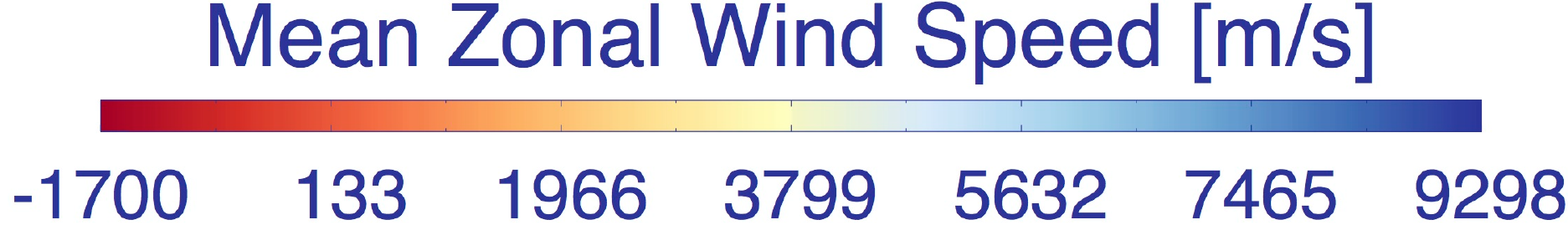}
\caption{The longitudinally averaged east-west (zonal) wind, as a function of latitude and pressure (the vertical coordinate) for all twelve rotational models. The tidally synchronous case is outlined in black. Note: color scale for these models is different than the color scale for the zonal wind maps presented in Figure \ref{fig:ZWM}. For the sake of uniformity, they are all on the same scale. The standard eastward equatorial jet seen in the synchronous rotation case becomes broader and then disrupted at slower rotation rates. At faster rotation rates it narrows and additional jets form at higher latitudes.}
\label{fig:appzonal}
\end{figure}

\section{Line-of-sight velocity maps} \label{sec:applos}
\begin{figure}[H]
\centering
\includegraphics[width=0.8\textwidth, keepaspectratio]{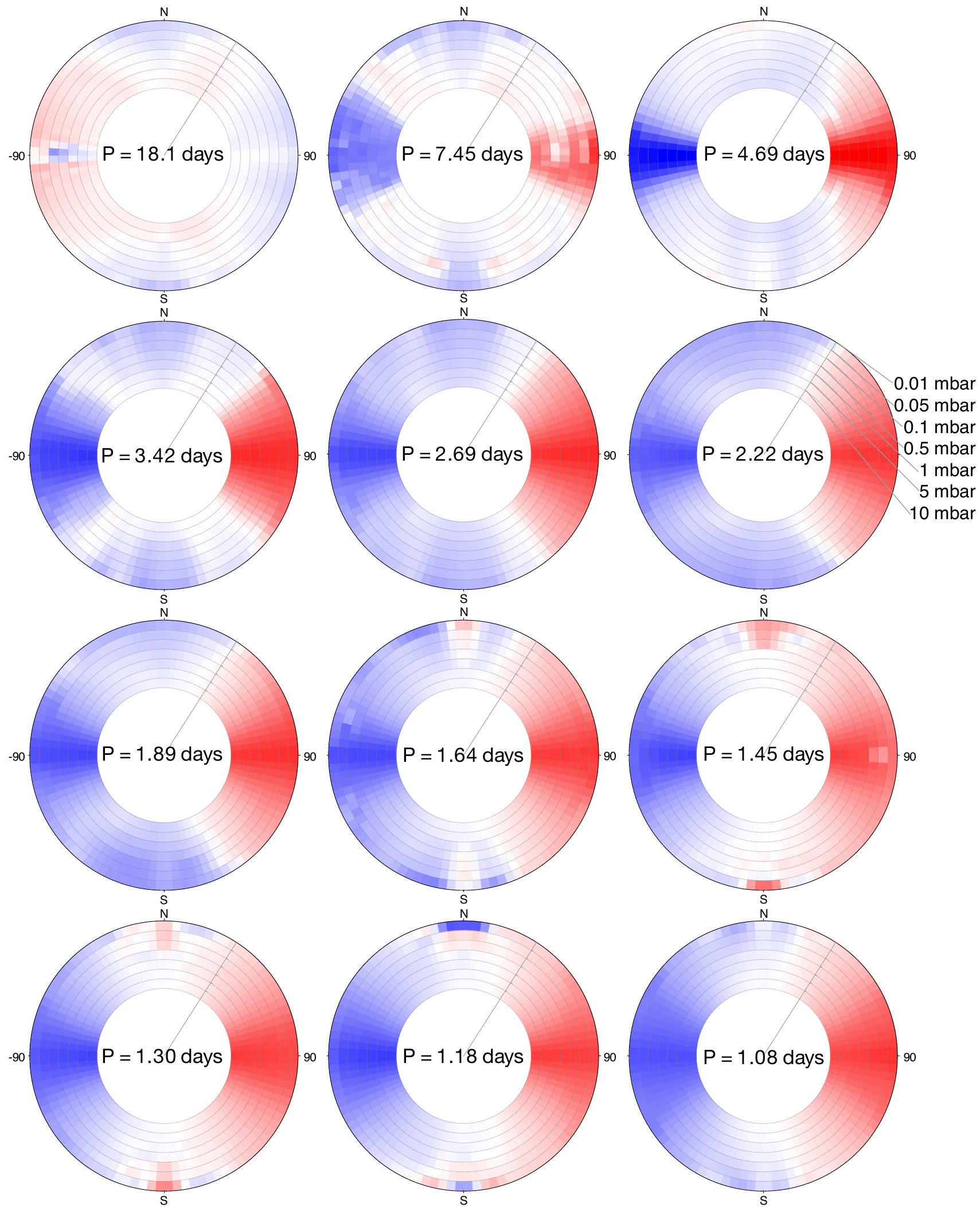}
\includegraphics[width=0.3\textwidth, keepaspectratio]{figure4_colorbar.pdf}
\caption{Cross-sections along the planet’s terminator, showing the line-of-sight velocities toward an observer during transit, for all twelve rotation models including the contributions from both winds and rotation. The pressure levels are all the same as the ones labeled in the tidally synchronous case. Even the most slowly rotating models have significant red- and blue-shifts at the terminator, due the presence of strong winds blowing east around the equator. These strong winds contribute to the broadening of and additional peaks in the cross-correlation functions (shown in Appendix \ref{sec:appccf}). Note: the atmosphere is not to scale with the size of the planet.}
\label{fig:applos}
\end{figure}

\section{Center cross-correlation functions}\label{sec:appccf}
\begin{figure}[H]
\centering
\includegraphics[width=0.9\textwidth, keepaspectratio]{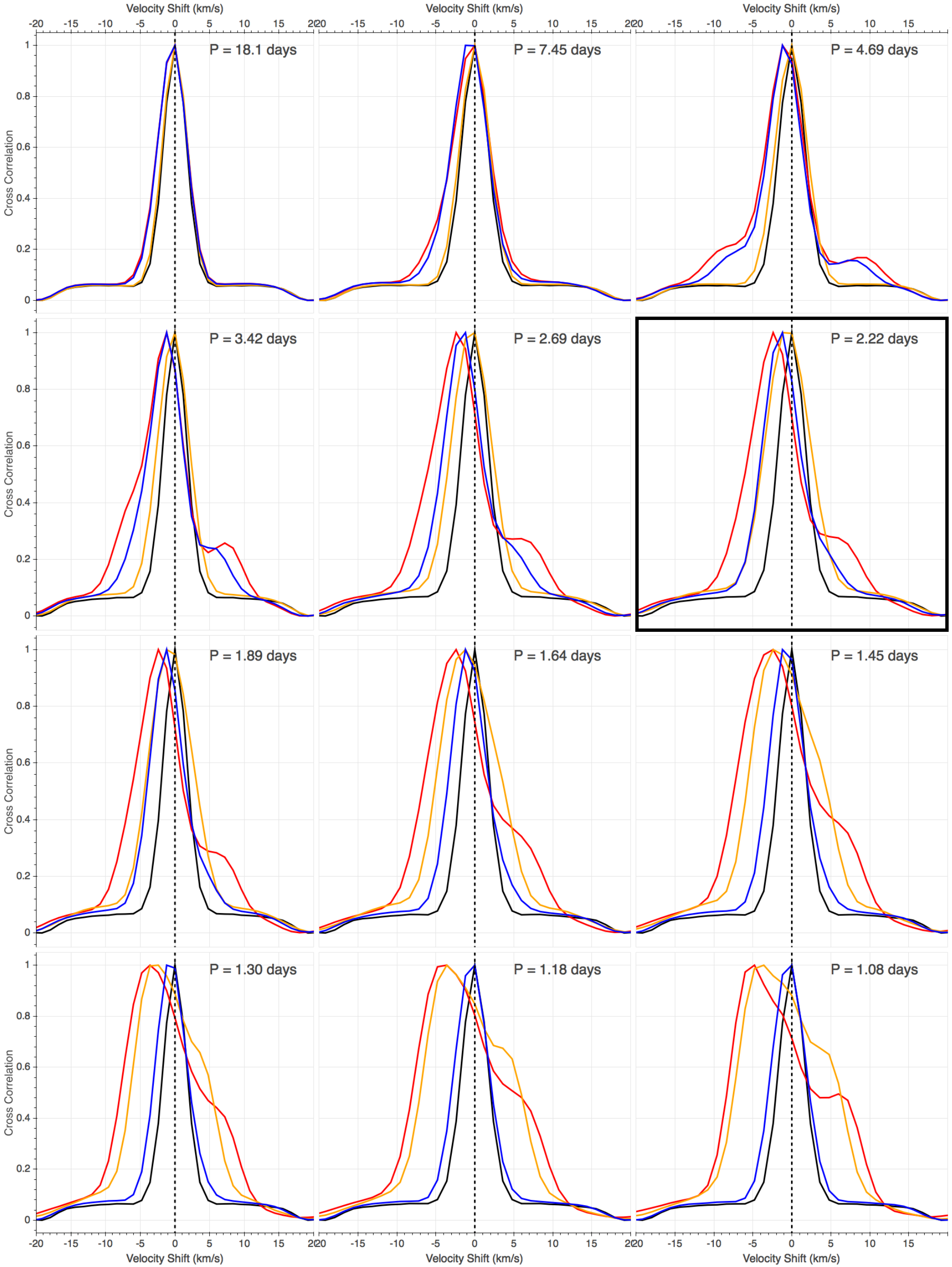}
\includegraphics[width=0.4\textwidth, keepaspectratio]{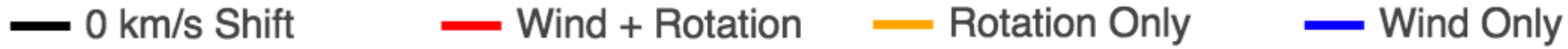}
\caption{Cross correlation functions at center transit for the twelve rotation models. Each model's rotation-only (yellow), wind-only (blue), and combined winds and rotation (red) transmission spectra were cross-correlated with the non-moving case in order to determine the velocity contribution of each process. The black line shows the non-moving case cross-correlated with itself as a reference point. Notably, winds are able to significantly broaden the line profiles of the most slowly rotating models, while in all models they contribute toward blue-shifting the net Doppler signature.}
\label{fig:appccf}
\end{figure}

\bibliographystyle{aasjournal}

\end{document}